\documentclass[lettersize,journal]{IEEEtran}
\usepackage{amsmath,amsfonts}
\usepackage{amssymb}
\usepackage{physics,amsmath}
\usepackage{algorithm}
\usepackage{algpseudocode}
\usepackage{array}
\usepackage{bm}

\usepackage{booktabs}
\usepackage{subcaption}
\usepackage{mathtools}
\usepackage{xcolor}
\usepackage{tabularx}
\usepackage{tikz}
\usetikzlibrary{chains}
\usepackage{makecell}
\usepackage{textcomp}
\usepackage{stfloats}
\usepackage{lipsum}  
\usepackage{bm}
\usepackage{url}
\usepackage{verbatim}
\usepackage{graphicx}
\usepackage{cite}
\usepackage{pdfpages}
\usepackage{tikz-3dplot}
\usepackage{relsize}
\usetikzlibrary{positioning, arrows.meta}
\usepackage{xcolor}
\usepackage[hidelinks]{hyperref}
\usepackage{siunitx} 
\usepackage{array} 

\usepackage{tikz}
\usetikzlibrary{3d}
\usetikzlibrary{arrows.meta} 

\definecolor{mycustomcolor1}{rgb}{0.6627, 0.1412, 0.1255} 
\definecolor{mycustomcolor2}{rgb}{0.8196, 0.5686, 0.2431}
\definecolor{mycustomcolor3}{rgb}{0.0549, 0.1922, 0.3765}
\definecolor{mycustomcolor4}{rgb}{0.0745, 0.3255, 0.5647}
\definecolor{mycustomcolor5}{rgb}{0.7882, 0.3490, 0.3725} 
\definecolor{mycustomcolor6}{rgb}{0.2157, 0.1843, 0.1843}
\definecolor{mycustomcolor7}{rgb}{0.9843, 0.8078, 0.2275}
\definecolor{mycustomcolor8}{rgb}{0.8784, 0.8157, 0.7216}
\newcommand{\Heaviside}{\text{H}}

\newcolumntype{C}[1]{>{\centering\arraybackslash}m{#1}}
\newcolumntype{L}[1]{>{\centering\arraybackslash}m{#1}}

\usepackage{courier} 
\usepackage{algorithm} 
\usepackage{algpseudocode} 

\algrenewcommand\alglinenumber[1]{\small\ttfamily\textcolor{black}{#1}}
\algrenewcommand\algorithmicrequire{\textbf{\small\ttfamily Input:}}
\algrenewcommand\algorithmicensure{\textbf{\small\ttfamily Output:}}
\algrenewcommand\algorithmiccomment[1]{\hfill\#\ \eqparbox{COMMENT}{\small\ttfamily #1}}

\begin{document}

\newcommand{\orcidiconAbk}{\href{https://orcid.org/0009-0006-1187-7782}{\includegraphics[scale=0.1]{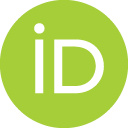}}}

\newcommand{\orcidiconOba}{\href{https://orcid.org/0000-0003-2523-3858}{\includegraphics[scale=0.1]{figures/orcidID128.png}}}

\title{ End-to-End Mathematical Modeling of Stress Communication Between Plants}

\author{Ahmet B. Kilic\orcidiconAbk,~\IEEEmembership{Student Member,~IEEE}, 
        and Ozgur B. Akan\orcidiconOba,~\IEEEmembership{Fellow,~IEEE}                
        \thanks{The authors are with the Center for neXt-generation Communications (CXC), Department of Electrical and Electronics Engineering, Koç University, Istanbul, Turkey (e-mail: \{ahmetkilic20, akan\}@ku.edu.tr).}
        \thanks{Ozgur B. Akan is also with the Internet of Everything (IoE) Group, Electrical Engineering Division, Department of Engineering, University of Cambridge, Cambridge, CB3 0FA, UK (email: oba21@cam.ac.uk).} 
	    \thanks{This work was supported in part by the AXA Research Fund (AXA Chair for Internet of Everything at Ko\c{c} University).}
}


	
\maketitle
\begin{abstract}
Molecular Communication (MC) is an important communication paradigm found in nature. Odor-based Molecular Communication (OMC) is a specific type of MC with promising potential and a wide range of applications. In this paper, we examine OMC communication between plants in the context of stress communication. Specifically, we explore how plants use Biological Volatile Organic Compounds (BVOCs) to convey information about the stresses they are experiencing to neighboring plants. We constructed an end-to-end mathematical model that discovers the underlying physical and biological phenomena affecting stress communication. To the best of our knowledge, this is the first study to model this end-to-end stress communication. We  numerically analyzed our system under different scenarios using MATLAB. Using experimental data from the literature, we demonstrated that continuous gene regulation can approximate BVOC emissions in plants under different stress conditions. Consequently, we applied this model to these stressors and plants to accurately approximate BVOC emissions. We also investigated a modulation method that plants use to send their messages, namely Ratio Shift Keying. Upon analyzing this method, we found that it benefits plants by both enabling a multiple access channel and preventing competitor plants from obtaining the information.

\end{abstract}

\section{Introduction}
\IEEEPARstart{M}{olecular} communication (MC) is an important communication paradigm that employs molecules to transmit information \cite{Kilinc2013,7747509}. Examples of MC can be found in nature and have existed for billions of years \cite{7747509}. This type of communication occurs between cells over short distances \cite{mcsurvey}, whereas pheromones facilitate communication over longer ranges \cite{longrangeMC}.

A special type of MC is called odor-based molecular communication (OMC), where pheromones or odors are used to transmit information \cite{bilgen2024odorperceptualshiftkeying}. OMC systems can benefit various fields \cite{bilgen2024odorperceptualshiftkeying}. Potential applications for these systems include the biomedical, industrial, and consumer goods industries, as well as telecommunications, information and communications technology, and future internet applications \cite{longrangeMC,aktas2023odorbasedmolecularcommunicationsstateoftheart,bilgen2024odorperceptualshiftkeying}. Therefore, to realize the potential of OMC systems, OMC systems in nature should be investigated.

In nature, OMC can be observed as plant-animal communication, plant-plant communication, animal-animal communication, and animal-bacteria-animal communication \cite{aktas2023odorbasedmolecularcommunicationsstateoftheart}. Among these communication types, inter-plant communication can be understood through OMC systems, providing insights into the plant world \cite{bilgen2024mycorrhizal,babar2024sustainable,bilgen2024odorperceptualshiftkeying,aktas2023odorbasedmolecularcommunicationsstateoftheart}. Plants are exposed to both biotic and abiotic stresses, which harm them and cause global economic damage \cite{Midzi2022}. Biotic stresses include insects, bacteria, viruses, fungi, nematodes, arachnids, and weeds, while abiotic stresses include drought, extreme temperatures, and salinity \cite{Midzi2022}. When exposed to these stresses, plants release Biological Volatile Organic Compounds (BVOCs) to warn neighboring plants of upcoming threats \cite{Midzi2022}. The effect of different stressors on the release of BVOCs has been demonstrated experimentally in various studies \cite{ruan2019jasmonic,karban2000communication,wang2021function,cheong2003methyl,beauchamp2005ozone,behnke2009rnai,pazouki2016mono,li2017ozone,acton2018effect,kanagendran2018differential,mengistu2014contrasting,yli2016herbivory,brilli2011detection,erb2015indole,portillo2015emission,faiola2015impacts,jiang2017methyl}. This is known as stress communication between plants, where the plant releasing the stress signal is called the transmitter plant, and the others are receiver plants. Understanding this communication can help prevent losses in agriculture and contribute to sustainable agricultural practices \cite{Midzi2022}. Manipulating plant stress signaling pathways could lead to novel strategies for sustainable crop protection by enhancing plant defenses \cite{Midzi2022}. Plant communication through BVOCs is not limited to stress communication; plants also interact with other plants and animals through BVOCs in various contexts \cite{aktas2023odorbasedmolecularcommunicationsstateoftheart}. Therefore, understanding stress communication can also aid in exploring other types of communication in plants.

 Although the exchange of information between plants has been studied in the field of MC \cite{10208114,9399099,Bige2016}, to the best of our knowledge, there is no mathematical model that explains end-to-end stress communication between plants. Therefore, this study investigates stress communication between plants while establishing the system components fundamental to a conventional communication method within this context.

The rest of this paper is structured as follows. Section II explains the mathematical modeling and defines the system components of stress communication. Section III examines how system parameters affect stress communication and validates the model of stress-induced emission. In Section IV, a modulation technique for stress communication, observed in nature, that enables encryption and multi-user channels is discussed. Finally, conclusions are drawn in Section V.

\section{Mathematical Modeling of Stress Communication}
\label{sec:mathmodel}

\begin{figure*}[t]
    \centering
    \includegraphics[width=\linewidth, height=0.5\linewidth]{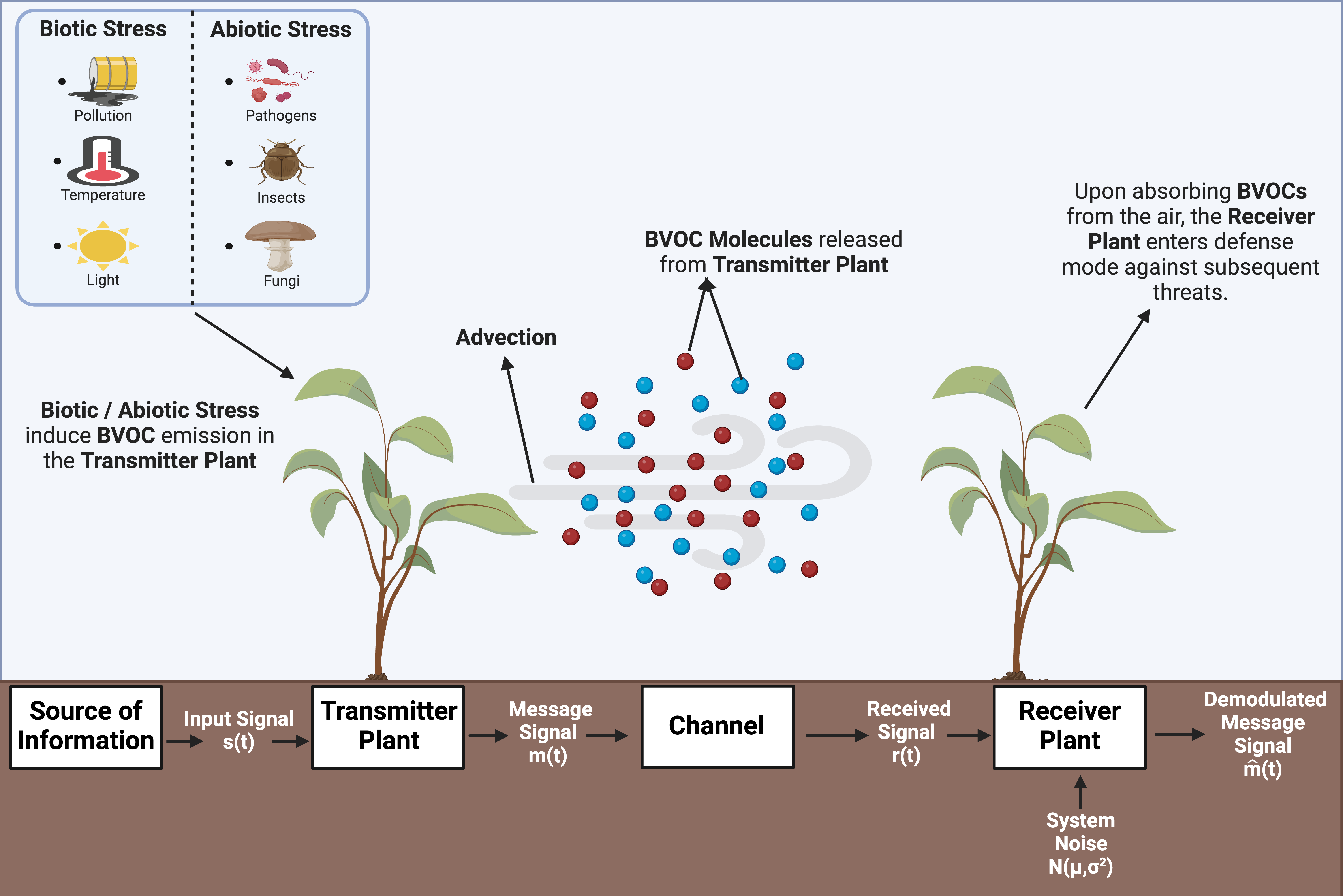}
    \caption{Overall scheme of end-to-end stress communication between plants \cite{fig_overall}.}
    \label{fig:combined}
\end{figure*}

In this section, end-to-end stress communication between plants is mathematically modeled and analyzed using communication theory. This communication is examined in three parts, namely the transmitter plant, the open-air communication channel, and the receiver plant. First, stress as the information source and the resulting biogenic volatile organic compounds as message signals are modeled in the transmitter plants. Then, the propagation of these BVOCs in the open air is modeled and investigated using turbulent diffusion formulas. Lastly, the uptake of these BVOCs by the receiver plant is examined, including an analysis of the decoding process of the message signals. The overall scheme of this communication can be found in Fig. \ref{fig:combined}.

\subsection{Transmitter Plant}
In conventional communication systems, it is necessary to express the source of information and the resulting message signal that is encoded at the transmitter. In the context of stress communication between plants, the source of information is the external stress factors affecting the transmitter plant. Additionally, the message signal is the BVOCs emission resulting from the stress. After modeling the transmitter plant from the OMC perspective, this section models the external stress on the transmitter plant and the resulting BVOC emission.

\subsubsection{OMC Modeling of Transmitter Plant}
\label{sec:OMCtransmitter}
An OMC transmitter consists of odor containers that store odor used in communication and a molecule delivery unit by which the transmitter can precisely control the release of molecules through diffusion, heat, or active pumping mechanisms \cite{aktas2023odorbasedmolecularcommunicationsstateoftheart}. Plants can also be investigated from the perspective of OMC, and the components of an OMC transmitter can be observed in plants as well. Plants have both specific and non-specific storage compartments for BVOCs in their leaves \cite{niinemets2004physiological}. These compartments can be considered analogous to odor containers in the OMC system. In these compartments, BVOCs are found in a variety of forms, such as liquid, lipid, and gaseous \cite{niinemets2004physiological}. From these compartments, BVOCs are transported to the leaf surface to be released into the air. On the leaf surface, BVOCs are released through small pores called stomata or a waxy layer called the cuticle \cite{niinemets2004physiological}. The release of BVOCs from small pores can be modeled as a diffusion mechanism in an OMC transmitter, where odor molecules move from higher concentration to lower. On the other hand, the release of BVOCs from stomata can be modeled as an active pumping mechanism in an OMC transmitter, where stored odor molecules are released by external pressure applied by the transmitter. An OMC transmitter releases the message signal given an input signal from the source of information. In the context of stress communication, the source of information is the stress on the plant.

\subsubsection{Source of Information in Transmitter Plant}
\label{sec:sourceofinfo}
Given source of information, the input signal must be identified. For most of the stressors, input signal can be identified as the immediate effect of the stressor on the plant. For example, in the case of mechanical damage, input signal is the length of the damage on the subject plant. Similarly, for stress caused by insects, input signal is the area of leaves eaten. However, finding this immediate effect of the stressor may not be straightforward. For example, in the case of heat stress, input signal is considered to be temperature, as the immediate effect of heat on the plant is not clear. The source of information can be both abiotic and biotic stress. Although their effects on regulating BVOC emission can differ significantly, assuming them as upstream regulators, input signal can be approximated by a polynomial of degree $n$ as
\begin{equation}
\label{eq:stressfactor}
    s(t) \approx  a_0 + a_1t + a_2t^2 + ... + a_nt^n,
\end{equation}
where the polynomial fit is utilized as an estimate of an underlying ``true" expression profile, which is masked by experimental errors, and coefficients $\{a_0,.., a_n\}$ are derived from the experimental gene expression profile using a least squares minimization approach \cite{vu2007nonlinear}. Additionally, it is assumed that the experimental error's weight is constant across all measurement locations.

\subsubsection{Generating Message Signal in Transmitter Plant}
\label{sec:message}
Stress-induced transmitter plant produces and emits BVOCs to transport information to the receiver plant. To model the emission of BVOCs from the plants, production rate of the BVOCs and the internal process of the plants releasing them should be investigated. 
 
Upon production, BVOCs are emitted through two primary pathways: they are either released from internal storage compartments within the leaves or emitted immediately \cite{harley2013stoma}. For the former, various physiological and physicochemical controls affect the release from these compartments \cite{harley2013stoma}. To develop a broadly applicable model, the effect of storage compartments on the emission rate is disregarded, and only the immediate emission pathway is considered. This approach aligns with the fact that the effect of storage compartments on the emission rate is temporary when integrated over time \cite{harley2013stoma}. Furthermore, following production, the majority of induced BVOCs are released directly rather than from storage compartments \cite{grote2019modeling}.

To link the emission rate and production rate of BVOCs, stomatal effects on the emission of produced BVOCs should be investigated. According to \cite{harley2013stoma}, stomata exert varying degrees of control over BVOC emissions. However, when predicting emission rates for incorporation into chemistry-transport models, emissions are primarily influenced by the production rate, with stomatal effects being less significant \cite{harley2013stoma}. Therefore, the effect of stomata on the emission rate is not included. This exclusion also allows for the development of a broadly applicable model, as stomatal effects vary among plants. Therefore, henceforth in this paper, the emission rate of BVOCs is assumed to be equivalent to their production rate.

The production rate, \textbf{I}, of BVOCs can be modeled using nonlinear differential equation model for quantification of transcriptional regulation that is proposed in \cite{vu2007nonlinear}. In this model, given a regulator, the effects of the regulator on gene activity changes over time are modeled. In the context of stress communication, gene activity changes over time correspond to the production rate of BVOCs, while the regulator represents the input signal in (\ref{eq:stressfactor}). Gene activity change over time, \(\frac{\partial g(t)}{\partial t}\), can be expressed as
\begin{equation}
\label{prod:single}
    I(t) = \frac{\partial g(t)}{\partial t} = \frac{\nu_{max}}{1+\exp(-ws(t)+c)} - k_d \cdot g(t),
\end{equation}
where \(v_{max}\) represents the highest possible rate at which a gene can produce its product, \(k_d\) denotes the rate at which this gene product degrades over time, \(g(t)\) indicates the quantity of the gene product present at time \(t\), \(s(t)\) represents input signal as defined in  (\ref{eq:stressfactor}), \(w\) is the regular constant determining gene operation, and \(c\) accounts for any delay in the gene's response to external signals, specifically describing the time it takes for transcription to commence after receiving a signal.

Although this production rate can be used to model the emission of BVOCs resulting from stress, the message signal conveying the information of the stress to the receiver plant should also be clearly defined. BVOCs are produced and released by plants regardless of the stress on the plant \cite{preston2001methyl}. This type of emission is called ordinary constitutive emission and occurs at much lower rates compared to production induced by stress \cite{preston2001methyl}. For a BVOC to be considered a message signal between plants, its release quantity in response to stress should significantly differ from the quantity of ordinary constitutive emissions of the same BVOC in the transmitter plant \cite{preston2001methyl}. Hence, variations in the amount of BVOCs absorbed by the receiver plant lead to a corresponding response to the message signal \cite{preston2001methyl}. Therefore, similar to the Concentration Shift Keying (CSK) method in molecular communication systems \cite{cskmsk}, it can be suggested that information about stress is encoded in the amount of BVOCs emitted. This is supported by the fact that there is a clear link between the level of stress a plant experiences and the quantity of chemicals it releases \cite{copolovici2011volatile}. Therefore, to clearly distinguish the message signal from constitutive emissions and to encode the severity of stress in the message signal, the message signal in stress communication can be defined as
\begin{equation}
m(t) = \left( \Heaviside(t - \tau_b) - \Heaviside(t - \tau_e) \right) \int_{\tau_b}^{\tau_e} I(t) \, dt,    
\end{equation}
where $\Heaviside(t)$ is Heaviside step function, $I(t)$ is the production rate defined in (\ref{prod:single}), $\tau_b$ is the time when the production rate differs from the constitutive emission rate, and $\tau_e$ is the time when the production rate returns to the constitutive emission rate.

\subsection{Open-air Communication Channel}
\label{sec:openair}
In this section, the mathematical modeling of BVOCs propagating in the air is investigated. In the context of this study, the communication channel in stress communication is a 3-D advection channel environment. The propagation of information molecules in the channel is driven by turbulent diffusion and advection caused by wind. After BVOCs are emitted from the transmitter plant, they propagate through the air to reach the receiver plant. To calculate the amount of reaching BVOCs, we need to know the concentration of the emitted BVOCs at the location in the channel where the receiver plant is situated. The concentration of molecules in the presence of diffusion and advection is formulated as
\begin{equation}
    \frac{\partial C}{\partial t} = D \cdot \nabla^2 C - \nabla \cdot (\boldsymbol{\vec{u}} C) + T,
    \label{eq:pde}
\end{equation}
where T denotes the source term, $\boldsymbol{\vec{u}}$ represents the flow rate vector, $D$ denotes the diffusion coefficient of odor molecules in the given medium,  and $C$ is the concentration function over space and time. 

To calculate concentration of BVOCs at the location in the channel where the receiver plant is located, (\ref{eq:pde}) should first be solved by setting up initial assumptions to obtain a closed-form solution for cases of turbulent diffusion. First, the channel is considered to be an isotropic turbulent field, where the eddy diffusivities are only dependent on the downwind distance $x$.  Second, at the plane $z = 0$, a flat surface is taken to be the ground. Third, the flow rate vector, $\boldsymbol{\vec{u}}$, which represents the wind in nature, is also assumed to be constant and aligned with the positive x axis. Hereby, $\boldsymbol{\vec{u}}$ can be represented as $\boldsymbol{\vec{u}} = (u,0,0)$ for constant $u$. Fourth, since wind diffusion is greatly significant compared to turbulent diffusion in x-direction, turbulence diffusion in this direction is disregarded. This assumption, as discussed in \cite{Bige2016}, is supported by the literature \cite{Roberts2000, Pasquill1968, Marsden1993}. Lastly, the emission of BVOCs is assumed to be instantaneous at $t = 0$. Although BVOC production occurs over time, we assume all of the produced BVOCs are released at  $t = 0$. This is a common assumption, when release time is small compared to the travel time of BVOCs \cite{silva2013puff}. Using (\ref{eq:pde}) and the initial assumptions, the concentration of BVOC molecules is formulated in \cite{Bige2016} as
\begin{equation}
    \begin{aligned}
    C(x,y,z,t) = \frac{M}{8(\pi k)^{3/2}} &e^{\frac{-(x-ut)^2-y^2}{4k}} \\&  \cdot\left[ e^{\frac{-(z-h)^2}{4k}} + e^{\frac{-(z+h)^2}{4k}} \right],
    \end{aligned}
\label{eq:cform}
\end{equation}
where $h$ is the vertical distance of the source from the origin in the $z$ direction and $M$, the total mass of BVOCs released, represents the amount of message signal in \si{\kilo\gram}. The variable $k$ is defined as 
\begin{equation}
    k = \frac{1}{u}\int_0^xD(\eta)\,d\eta,
\end{equation}
to eliminate analytical difficulties caused by eddy diffusivities, $D$ \cite{Bige2016}.

\subsection{Receiver Plant}
To conclude the mathematical modeling of stress communication between plants, the uptake of BVOCs at the leaves of the receiver should also be modeled. BVOCs at the reception area of the leaves are absorbed by various methods \cite{trapp2007fruit}. Diffusion through cuticles and stomata are some of the primary ways leaves uptake BVOCs \cite{trapp2007fruit}. Analogously, this absorption process can be best modeled as an absorbing receiver in the context of OMC. Every molecule that comes into contact with an absorbing receiver is absorbed and broken down \cite{kuscutransmittersurvey}. Given the nature of BVOC uptake in plants, receiver plants can be modeled as absorbing receivers.

While modeling the reception in the OMC, redundant molecules from previous messages are included in the calculation because they create inter-symbol interference (ISI) at the receiver. However, since the main purpose of this paper is to analyze the communication of stress caused by a new stressor, it is unnecessary to incorporate the effects of previous stresses into the transport model. This is supported by the fact that emissions from a single stress can last from hours to days, and the effects of previous BVOCs will have diminished significantly over such long time intervals \cite{mengistu2014contrasting,copolovici2011volatile}. Hence, in the context of this paper, the memory of the OMC channel is disregarded, and emissions are assumed to be single-shot. Consequently, the fate of unabsorbed molecules is not investigated.

Both dynamic \cite{rein2011new} and steady-state \cite{trapp1995generic, trapp2007fruit} models are employed to mimic BVOC uptake in plant leaves. To simplify the modeling of stress communication between plants, the steady-state solution model suggested in \cite{trapp2007fruit} is employed. This model is consistent with the results of \cite{trapp1995generic}, given the same inputs \cite{trapp2007fruit}. Thus, it can be stated that the steady-state solution accurately represents BVOC uptake in the leaves. The mass of BVOCs in the leaves can be calculated by adding the uptaken BVOCs and translocation from the stem, then subtracting metabolism and loss to air \cite{trapp2007fruit}. In this paper, because our focus is on modeling the target BVOC in the receiving plant, the effects of metabolism and translocation from the stem are dismissed. This is also consistent with the model suggested in \cite{trapp1995generic}. Therefore, uptake of BVOCs at the receiver plants can be modeled as 
\begin{equation}
\frac{dC_L(t)}{dt} = \frac{P_L \cdot A_L}{K_{AW} \cdot M_L} \cdot C_{Air}(t) - \frac{1000 \cdot P_L \cdot A_L}{K_{LW} \cdot M_L} \cdot C_L(t), 
\label{eq:uptake}
\end{equation}
where \(C_L\) is the concentration of BVOC in the leaf, \(P_L\) is the permeability, \(A_L\) is the area of the leaf, \(C_{Air}\) is the concentration of BVOC in the air at the location of the receiver plant, \(M_L\) is the total mass of the leaf, \(K_{LW}\) is the partition coefficient between leaf and water, and \(K_{AW}\) is the partition coefficient between air and water. The concentration of BVOCs in the air at the location of the receiver plant, \(C_{Air}\), can be found as 
\begin{equation}
\label{eq:cair}
    C_{Air}(t) = C(x_r,y_r,z_r,t),
\end{equation}
where $(x_r,y_r,z_r)$ is the location of the receiver plant. In a communication system, the received signal is the information-carrying signal that reaches the receiver after being transmitted through a channel. Identifying the received signal can help assess channel performance and the effect of the channel on the message signal. Since the concentration of BVOCs in the air at the location of the receiver plant is already identified as \(C_{Air}(t)\), the received signal in the context of stress communication can be identified as
\begin{equation}
     r(t) = C_{Air}(t).
\end{equation}

After solving (\ref{eq:uptake}) using antidifferentiation \cite{adkins2012ordinary}, we obtain

\begin{equation}
C_L(\tau_r) =  \frac{P_L \cdot A_L}{K_{AW} \cdot M_L}\int_0^{\tau_r} C_{Air}(t) \,dt,
\end{equation}
where $\tau_r$ is the time at which the reception and demodulation of the message signal occur at the receiver plant. The value of $\tau_r$ can vary under different scenarios and constraints. However, it is assumed that $\tau_r$ is chosen such that $C_L(\tau_r)$ is maximized. Replacing $C_{Air}(t)$ using the expression in (\ref{eq:cform}) and  organizing the terms result in 

\begin{equation}
\begin{aligned} 
C_L(\tau_r) &=  \frac{P_L \cdot A_L}{K_{AW} \cdot M_L}  \frac{M}{8(\pi k)^{3/2}} \\&\left( e^{\frac{-(z_r-h)^2-y_r^2}{4k}} + e^{\frac{-(z_r+h)^2-y_r^2}{4k}} \right)   \int_0^{\tau_r} e^{\frac{-(x_r-ut)^2}{4k}}\,dt .
\end{aligned}
\end{equation}

After calculating integrals, we obtain
\begin{equation}
\begin{aligned} 
\label{eq:CLlast}
C_L(\tau_r) =  \frac{P_L \cdot A_L}{K_{AW} \cdot M_L} &\frac{M}{8\pi k u} \left( e^{\frac{-(z_r-h)^2-y_r^2}{4k}} + e^{\frac{-(z_r+h)^2-y_r^2}{4k}} \right) \\&  \cdot \left(\operatorname{erf}\left(\frac{u{\tau}_{r} - x_r}{2 \sqrt{k}}\right) + \operatorname{erf}\left(\frac{x_r}{2 \sqrt{k}}\right)\right),
\end{aligned}
\end{equation}
where 

\begin{equation}
\text{erf}(z) = \frac{2}{\sqrt{\pi}} \int_{0}^{z} e^{-t^2} \, dt.
\end{equation}

So far, the amount of BVOCs absorbed by the receiver plant is calculated based on the release from the transmitter plant, but the effects of environmental, biological, and chemical noise have not been included. In communication systems, noise must be considered to understand how it affects information transmission. In this system, environmental noise could come from factors like temperature changes and humidity. Chemical noise may include background BVOCs that interfere with the message molecules, or the breakdown of these molecules due to conditions like ozone pollution \cite{Acton2018}. Biological noise might result from processes inside the plant, such as its metabolism, which can affect how BVOCs are absorbed.

Including all these types of noise in the system can be challenging. A common approach in MC is to model the noise as Gaussian noise at the output of the channel \cite{Kilinc2013}. Hence, calculation of BVOC concentrations at the receiver plant can be found as
\begin{equation}
\begin{aligned}
C_{LN}(\tau_r) = C_L(\tau_r) + \mathcal{N}(\mu, \sigma^2),
\end{aligned}
\label{eq:cnoisy}
\end{equation}
where $\textbf{N}(\mu,\sigma^2)$ is the Gaussian random variable modeling the system noise.

In the context of stress communication between plants, it is important to clearly define the demodulated message at the receiver plant. The demodulated message represents the decoded information from the message signal received by the plant. Receiver plants learn about the stress experienced by the transmitter plant through an increase in the amount of BVOCs taken up in their leaves \cite{copolovici2011volatile}. Thus, similar to the message signal, the demodulated message can be defined by the quantity of BVOCs present in the receiver plant's leaves. Since the amount of BVOCs in the leaf of the receiver plant is given by (\ref{eq:CLlast}), the demodulated message signal can be simply expressed as 

\begin{equation}
\label{eq:demod}
    \hat{m}(t)  =  C_{LN}(\tau_r) \cdot \delta(t-\tau_r).
\end{equation}

\begin{figure}[t]
    \includegraphics[width = \linewidth, height = 4cm]{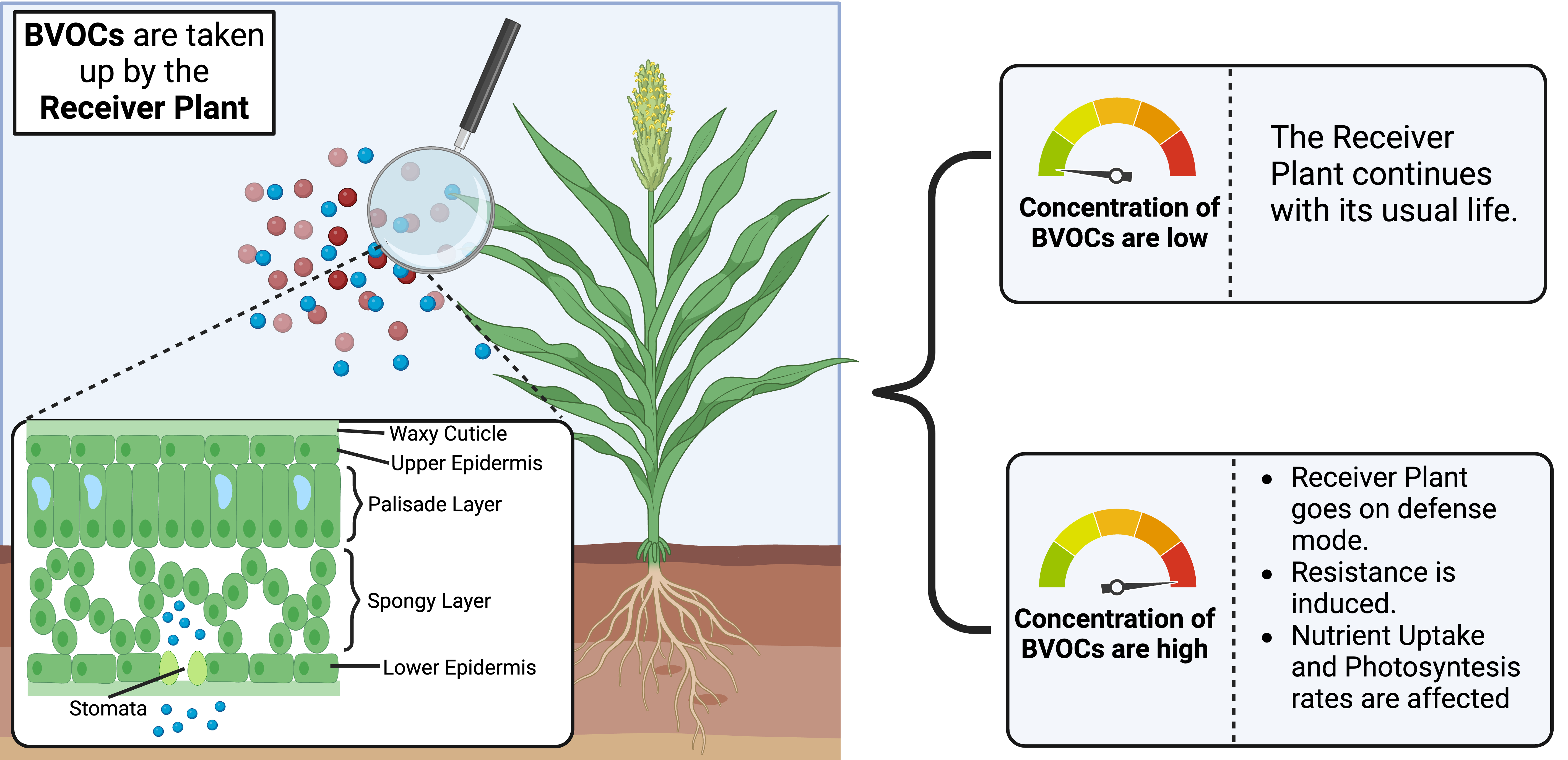 }
    \caption{Overall scheme of the demodulation process in the receiver plant \cite{fig_2}; plant responses are obtained from \cite{Midzi2022}.}
    \label{demod}
\end{figure}

The demodulated message signal can induce resistance to subsequent attacks and modulate primary and secondary metabolisms by affecting nutrient uptake and photosynthesis \cite{Midzi2022}. However, to prime the receiver plant, the concentration of BVOCs should be at an active level \cite{Midzi2022}. This phenomenon can be paralleled in MC systems as binary CSK, where the absence of information particles represents '0' and a concentration of information particles above a threshold represents '1'. In this analogy, if the concentration of BVOCs in the leaves of the receiver plant exceeds a threshold, the receiver plant receives the message '1', which puts the plant into a defense mode. Overall scheme of the demodulation process and how receiver plant reacts can be seen in Fig. \ref{demod}.

\section{Validation and Numerical Analysis}
So far, end-to-end stress communication between plants has been mathematically modeled and analyzed. In this section, the model will be further analyzed by validating the assumptions of the emission model, numerically investigating the mathematical model under different conditions.

First, the emission model used in Section \ref{sec:message} 
is validated using experimental data from various sources available in the literature. Then, the end-to-end mathematical model is analyzed to investigate the performance of the receiver plant in successfully demodulating the message signal by manipulating variables such as distance, wind speed, eddy diffusivity, and system noise.

\subsection{Validation of Stress-Induced Emissions}

 In \cite{grote2013leaf}, it proposed that induced BVOC emissions can be replicated through simulations using (\ref{prod:single}). In this study, the theory of continuous regulation of specific gene(s) over time was validated as a model for different stressors across various plants to accurately approximate BVOC emissions. Experimental data from the literature were utilized to fit the mathematical model and evaluate its accuracy. The experimental data were obtained from studies using Engauge Digitizer 12.1. 

The input signal in (\ref{eq:stressfactor}) should be determined first to approximate BVOC emissions. In the light of the discussion in Section \ref{sec:sourceofinfo}, input signal can be determined for different settings. After determining the input signal and obtaining experimental data, (\ref{eq:stressfactor}) should be fitted to these data points. For each experiment, the degree of the polynomial is selected based on the quantity of data points in the profile and the degree of fluctuations, as recommended in \cite{vu2007nonlinear}. Then, the coefficients of the polynomial are computed using a least squares minimization procedure, as also recommended in \cite{vu2007nonlinear}.

\begin{figure}[h!]
    \centering
    \includegraphics[height = 4cm]{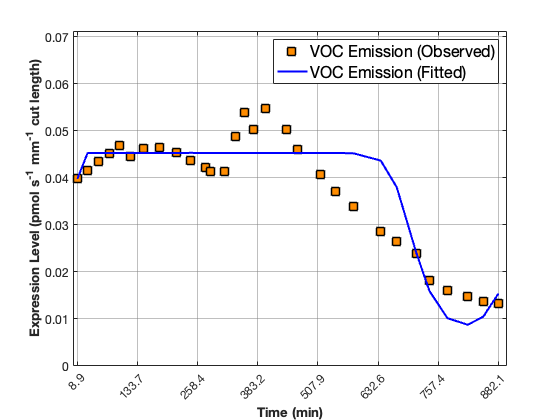}
    \caption{Approximation of induced Methanol emission with accuracy $r^2 = 0.7889$}
    \label{fig:methanol}
\end{figure}

\begin{figure*}[t]
    \centering
    \begin{subfigure}[b]{0.275\linewidth}
        \centering
    \includegraphics[width=\linewidth]{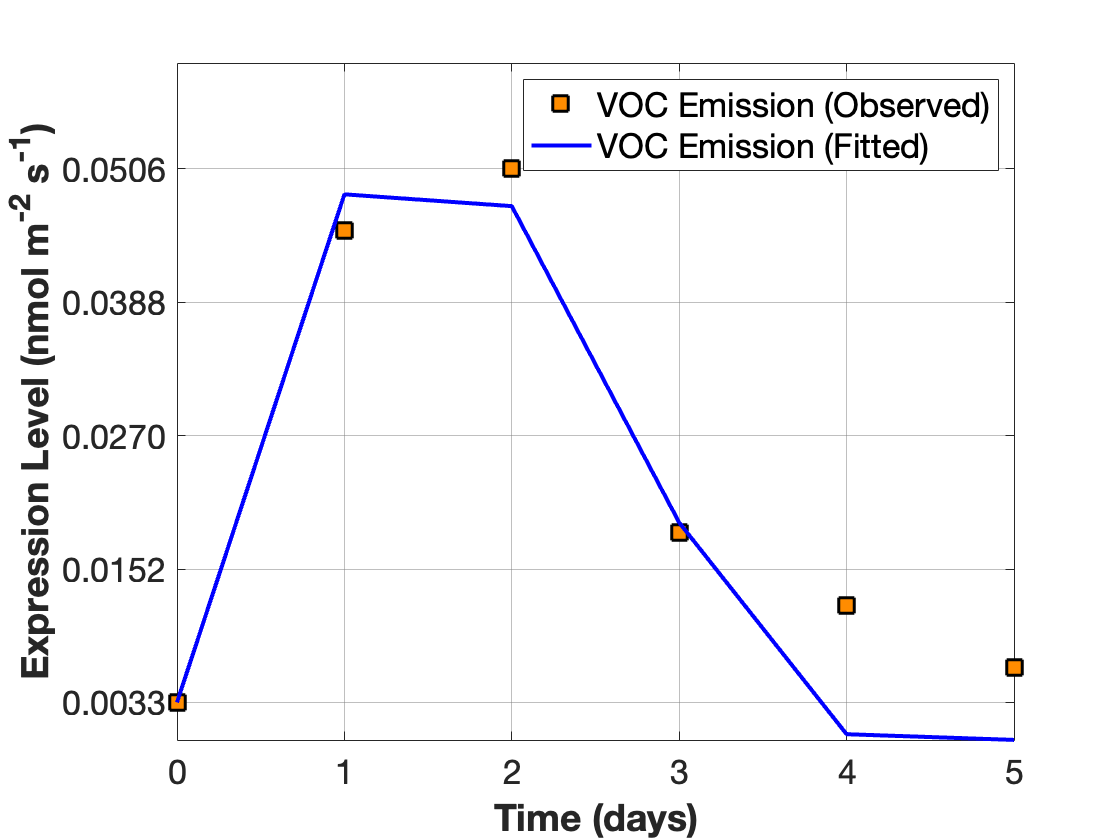}
        \caption{$r^2$: 0.9059}
        \label{fig:1}
    \end{subfigure}
    \hfill
    \begin{subfigure}[b]{0.275\linewidth}
        \centering
        \includegraphics[width=\linewidth]{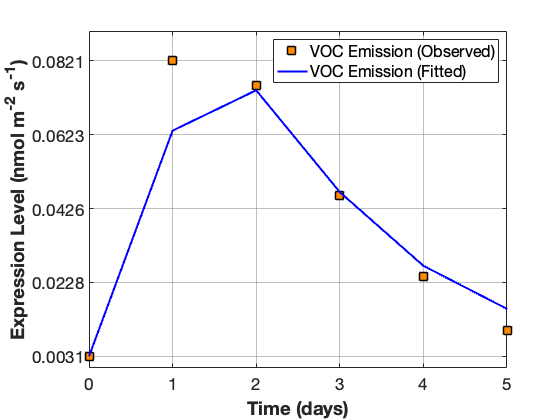}
        \caption{$r^2$: 0.9296}
        \label{fig:2}
    \end{subfigure}
    \hfill
    \begin{subfigure}[b]{0.275\linewidth}
        \centering
        \includegraphics[width=\linewidth]{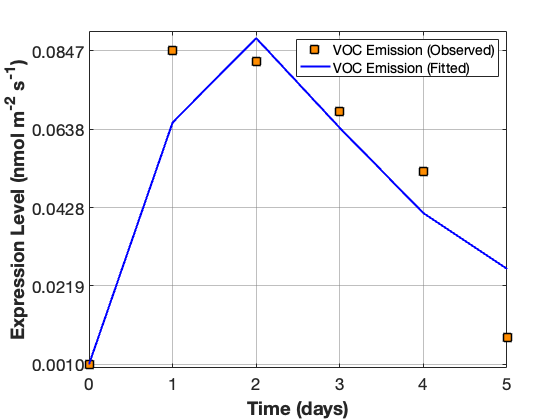}
        \caption{$r^2$: 0.8692}
        \label{fig:3}
    \end{subfigure}

    \vskip\baselineskip

    \begin{subfigure}[b]{0.275\linewidth}
        \centering
        \includegraphics[width=\linewidth]{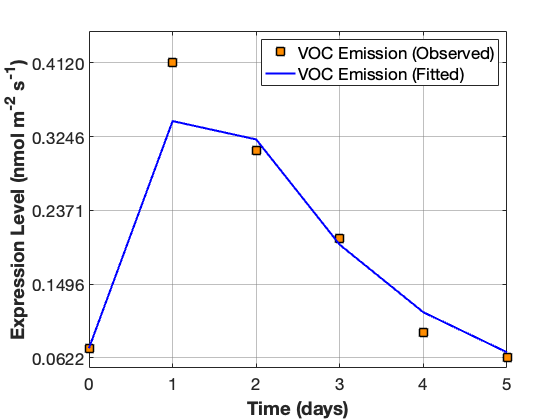}
        \caption{$r^2$: 0.9457}
        \label{fig:4}
    \end{subfigure}
    \hfill
    \begin{subfigure}[b]{0.275\linewidth}
        \centering
        \includegraphics[width=\linewidth]{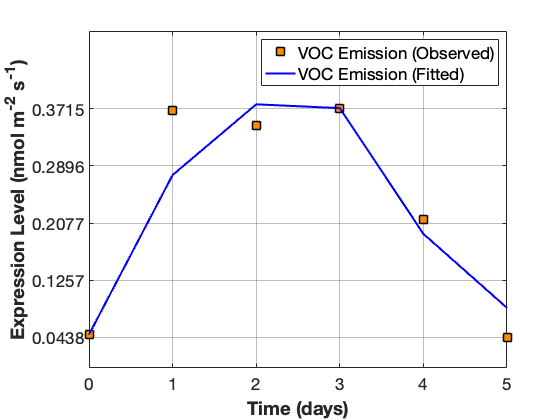}
        \caption{$r^2$: 0.9029}
        \label{fig:5}
    \end{subfigure}
    \hfill
    \begin{subfigure}[b]{0.275\linewidth}
        \centering
        \includegraphics[width=\linewidth]{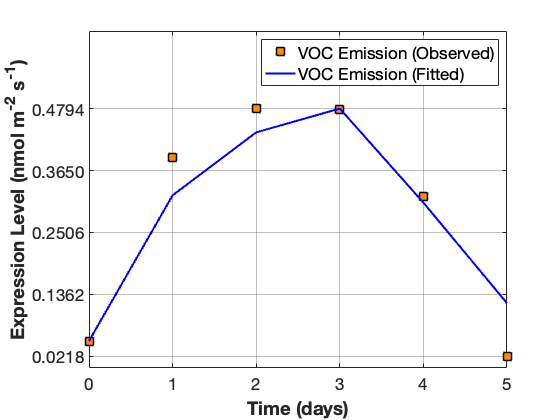}
        \caption{$r^2$: 0.9215}
        \label{fig:6}
    \end{subfigure}
    
    \caption{Accuracy for GLV and Monoterpene (GLVs on the upper row) for different herbivory levels (2, 4, 8).}
    \label{fig:overall}
\end{figure*}

After obtaining an expression approximating the stress effect on the plant, the BVOC emissions can be approximated using (\ref{prod:single}). Upon obtaining experimental data of BVOC emissions from the stressed plant, a least squares minimization procedure can be applied to determine the values of $w$, $c$ and $k_d$ that provide the closest approximation of BVOC emissions. The variable $v_{max}$ is simply calculated for each BVOC emission by selecting the highest emission rate during the emission period. After obtaining an approximation, the performance of the model is assessed by calculating the $r^2$ value between the actual and approximated emissions.

While validating the model, data from both biotic and abiotic stressors were used. In \cite{copolovici2011volatile}, the relationship between larval feeding and BVOC emissions is investigated. Using data from this study, our model achieved an accuracy of $r^2 > 0.9$ in modeling the quantitative relationship between stress intensity and BVOC emissions for both total lipoxygenase pathway (LOX) products and monoterpenes. Approximations for different herbivory levels for divverent BVOC emissions can be found in Fig. \ref{fig:1}-\ref{fig:6}.

Another study that is used for validating our model is \cite{brilli2011detection}. In this study, relationship between wounding of the plant and BVOC emission was studied.
The model's accuracy in simulating the quantitative link between stress intensity and methanol emission was 
$r^2 = 0.7889$ using the data from this study. Approximation for methanol emission can be found in Fig. \ref{fig:methanol}.

Lastly, data from \cite{pazouki2016mono} is used to validate our model. Unlike other studies that validate this model, this study examines the effect of an abiotic stressor, heat, on BVOC emissions. The accuracy of simulating the quantitative relationship between heat stress and total LOX emissions is $r^2 = 0.6232$. The lower accuracy, compared to biotic stressors, is due to the difficulty in modeling heat damage on leaves. For abiotic stressors, it is harder to quantify the extent of damage compared to biotic stressors, which can be quantified more easily.

\subsection{Numerical Analysis}
\label{sec:numanalyses}
In this part, the theoretical model presented in Section \ref{sec:mathmodel} will be numerically investigated. The effects of system parameters such as distance, wind speed, eddy diffusivity, noise intensity, and receiver threshold on the  $C_{LN}$  will be analyzed using MATLAB. Each analysis is aggregation of 10,000 trials. The list of analyses and the values of the system parameters for each analysis can be found in Table \ref{tab:analyses}. Additionally, constant system parameters that do not vary between analyses can be found in Table \ref{tab:constants}.

The mass ($M_L$) and area ($A_L$) of the leaf of are obtained from \cite{ganskopp1986estimating}, where calculations are made on \textit{Artemisia tridentata subsp. wyomingensis}. Permeability ($P_L$) and partition coefficient between air and water ($K_{AW}$) are obtained from \cite{trapp2007fruit}. The receiver's location is assumed to be along the direction of the wind, i.e., ($y_r = 0$), and at the same height as the transmitter plant, i.e., ($z_r = 1$) and ($h = 1$ \si{\meter}) . To ensure clarity and consistency, the motivation for choosing values for parameters that vary between different analyses is discussed in detail in each analysis.

\begin{table}[h]
  \centering
  \caption{Constant System Parameters} 
  \label{tab:constants}
  \begin{tabular}{|l|l|} 
    \hline
    \textbf{Parameter} & \textbf{Value} \\
    \hline
    $P_L$ &  $10^{-8}$ \si{\meter\per\second}\\
    $A_L$ & $25\cdot10^{-4}$ \si{\meter\squared} \\
    $M_L$ & $0.5\cdot10^{-3}$ \si{\kilo\gram} \\
    $K_{AW}$ & 10\\
    $(y_r,z_r)$ & $(0,1)$ \\
    $h$ & 1 \si{\meter} \\
    \hline
  \end{tabular}
\end{table}

\begin{table*}[t]
  \centering
  \caption{List of Analyses and System Parameters}
  \label{tab:analyses}

  \begin{tabular}{|L{2.5cm}|C{1.25cm}|C{2cm}|C{1.75cm}|C{2.75cm}|C{2cm}|C{2.2cm}|}
    \hline
    \textbf{Analysis Name} & \textbf{Distance ($x_r$)} & \textbf{Wind Speed (u)} & \textbf{System Noise} & \textbf{Receiver Threshold} &\textbf{Eddy Diffusivity (D)}&\textbf{Amount of Message Signal (M)}  \\
    \hline
    Distance Analysis (Fig. \ref{fig:distance} ) & 0 - 2 \si{\meter} & 25 \si{\meter\per\second} & $\mathcal{N}(\mu, \sigma^2)$  & 55\% $\cdot$ Max($C_{LN}$)  & 0.1 \si{\meter\squared\per\second} & $1.1\cdot10^{-9}$ \si{\kilo\gram}  \\
    \hline
    Distance - SNR Analysis (Fig. \ref{fig:distancesnr} ) & 0 - 2 \si{\meter} & 25 \si{\meter\per\second} & $\mathcal{N}(\mu, \sigma^2)$  & - & 0.1 \si{\meter\squared\per\second} &  $1.1\cdot10^{-9}$ \si{\kilo\gram}  \\
    \hline
    Distance - Delay Analysis (Fig. \ref{fig:distancedelay} ) & 0 - 2 \si{\meter} & 1 \si{\meter\per\second} & - & - & 0.1 \si{\meter\squared\per\second} & - \\
    \hline
    Distance - Mass Analysis (Fig. \ref{fig:distancemass} ) & 0 - 2 \si{\meter} & 25 \si{\meter\per\second} & $\mathcal{N}(\mu, \sigma^2)$  & - & 0.1 \si{\meter\squared\per\second} &  $[1.1,3.3,5.5,11] \cdot10^{-9} $ \si{\kilo\gram}  \\
    \hline
    Wind Speed Analysis (Fig. \ref{fig:speed} ) & 0 - 5.5 \si{\meter} &  [1, 25, 50, 100] \si{\meter\per\second} &  $\mathcal{N}(\mu, \sigma^2)$  &  - & 0.1 \si{\meter\squared\per\second}  &  $1.1\cdot10^{-9}$ \si{\kilo\gram} \\
    \hline
    Wind Speed - Delay Analysis (Fig. \ref{fig:speeddelay} ) & 10 \si{\meter} &  0 - 100 \si{\meter\per\second} &  $\mathcal{N}(\mu, \sigma^2)$  &  - & 0.1 \si{\meter\squared\per\second} & - \\
    \hline
    Eddy Diffusivity Analysis (Fig. \ref{fig:diffusivity}) & 0 - 1.5 \si{\meter}  & 25 \si{\meter\per\second} &  $\mathcal{N}(\mu, \sigma^2)$  & -    & [0.1,10,35,100] \si{\meter\squared\per\second}  &  $1.1\cdot10^{-9}$ \si{\kilo\gram}\\
    \hline
    Noise Intensity Analysis (Fig. \ref{fig:degradation} ) & 0 - 1.5 \si{\meter}  & 25 \si{\meter\per\second} &  [1,3,5,7,9]$\cdot \mu$ & - & 0.1 \si{\meter\squared\per\second} &  $1.1\cdot10^{-9}$ \si{\kilo\gram}  \\
    \hline
    Noise Intensity - SNR Analysis (Fig. \ref{fig:degradationsnr} ) & 0.75 \si{\meter}  & 25 \si{\meter\per\second} &  [$1-10$] $\mu$ & - & 0.1 \si{\meter\squared\per\second} &  $1.1\cdot10^{-9}$ \si{\kilo\gram}  \\
    \hline
    Receiver Threshold Analysis (Fig. \ref{fig:threshold} ) & 0 - 1 \si{\meter}  & 25 \si{\meter\per\second}&  $\mathcal{N}(\mu, \sigma^2)$ &  [2.5\%,5\%,12.5\%,50\%] $\cdot$ Max($C_{LN}$)   & 0.1 \si{\meter\squared\per\second} &  $1.1\cdot10^{-9}$ \si{\kilo\gram} \\
\hline
\end{tabular}
\end{table*}

\subsubsection{Distance Analysis}
\label{sec:distanceanalysis}
\begin{figure*}[t!]
    \centering
    \begin{subfigure}{0.245\linewidth}
        \includegraphics[width=\linewidth]{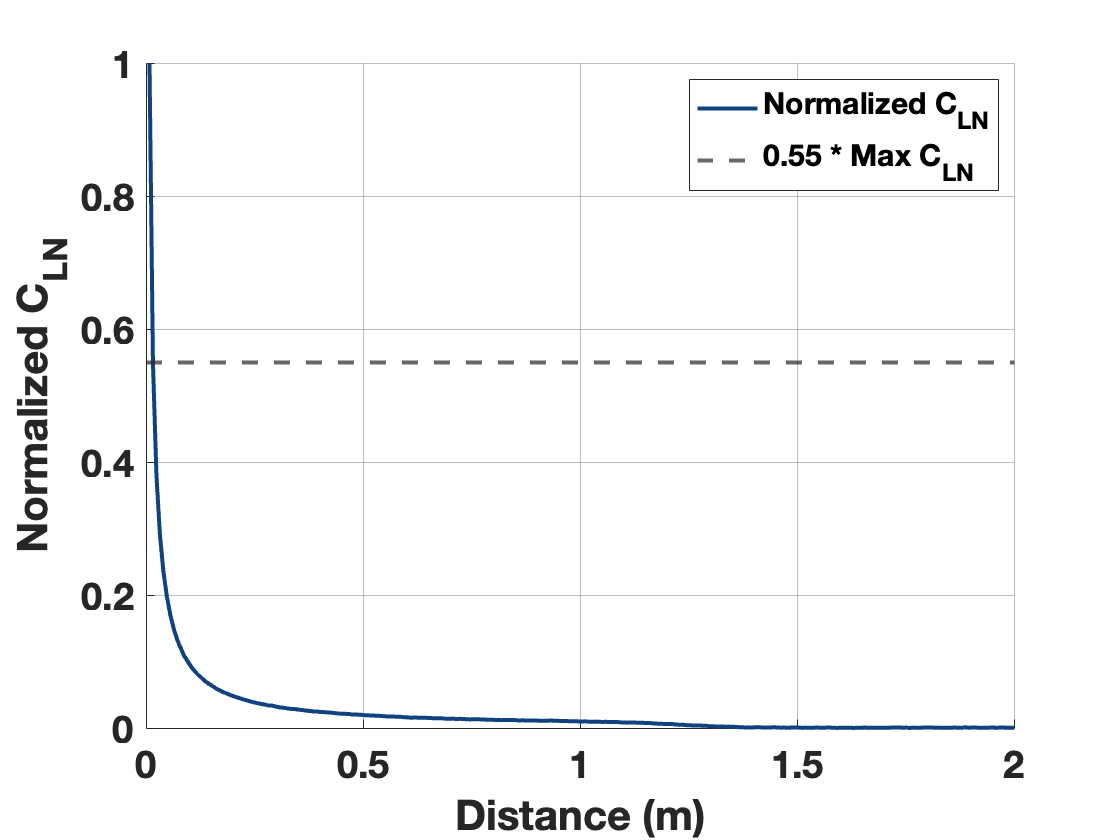}
        \caption{Distance Analysis}
        \label{fig:distance}
    \end{subfigure}
    \begin{subfigure}{0.245\linewidth}
        \includegraphics[width=\linewidth]{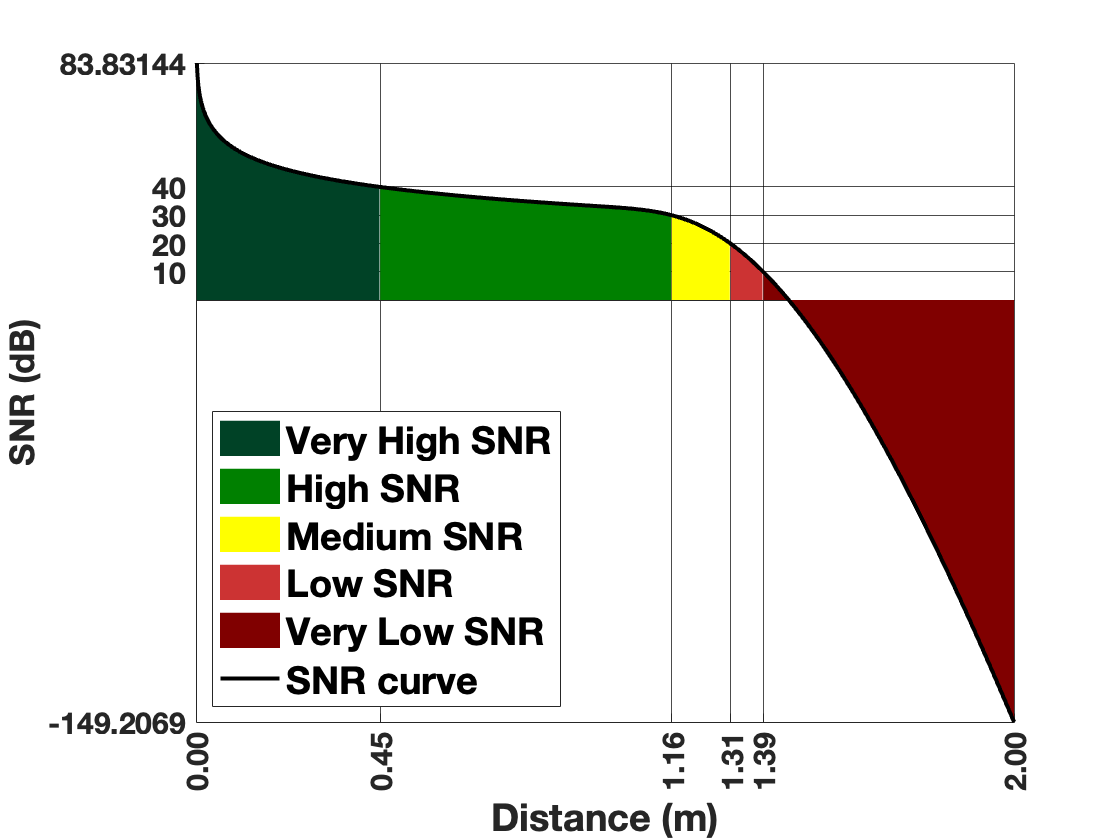}
        \caption{Distance - SNR Analysis}
        \label{fig:distancesnr}
    \end{subfigure}  
    \begin{subfigure}{0.245\linewidth}
        \includegraphics[width=\linewidth]{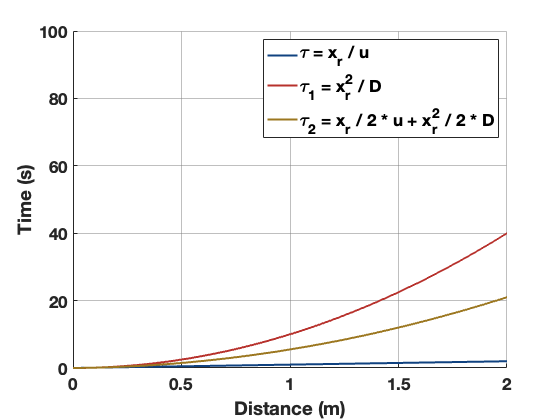}
        \caption{Distance - Delay Analysis}
        \label{fig:distancedelay}
    \end{subfigure}
    \begin{subfigure}{0.245\linewidth}
        \includegraphics[width=\linewidth]{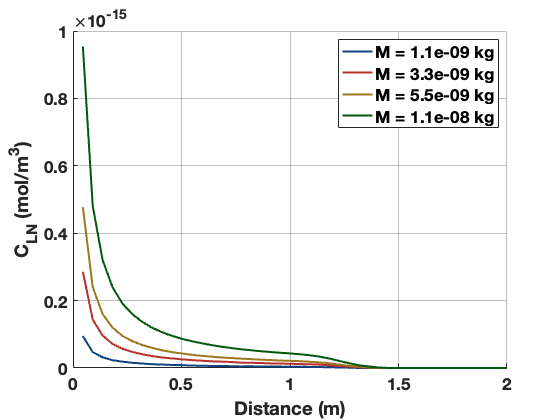}
        \caption{Distance - Mass Analysis}
        \label{fig:distancemass}
    \end{subfigure}   
    \caption{Distance Analyses}
    \label{fig:DistanceAnalyses}
\end{figure*}
In this analysis, the successful demodulation of the message signal in the receiver plant is investigated for distances between plants ranging from 0 to 2 \si{\meter}. These distances are chosen based on actual plant stress communication data. It has been observed that effective stress communication in nature can range up to 6 to 7 meters \cite{hirose2024theoretical}.

For the receiver plant to successfully demodulate the message signal, as discussed in Section \ref{sec:message}, the amount of BVOC uptake must vary significantly. Specifically, the BVOC concentration must exceed a threshold in the receiver plant to induce a response \cite{hirose2024theoretical}. This threshold value varies between plants. Additionally, this threshold can even change for the same plant over time if it has been induced earlier \cite{Midzi2022}. Therefore, in this analysis, the threshold value is assumed to be 55\% of the emitted BVOC. Further analysis on the effect of different threshold values on successful demodulation can be found in Section \ref{sec:thanalysis}. Given the threshold value, only receiver plants located between 0 - 0.1 \si{\meter} can successfully demodulate the message signal. For receiver plants located further than 0.1 \si{\meter}, the inability to obtain a sufficient concentration of BVOCs can be attributed to the dispersion of BVOCs outside the reception volume due to diffusion. 

Obtaining the information of distance on the concentration of BVOCs, another important performance metric can be investigated. Signal-to-noise ratio (SNR) can be calculated for distances between plants ranging from 0 to 2.5  \si{\meter}. SNR can be formulated as 
\begin{equation}
    SNR = \frac{P_{signal}}{P_{noise}},
\end{equation}
where $P_{signal}$ is the average power of the message signal and $P_{noise}$ is the average power of the background noise \cite{johnson2006signal}. The noise in the context of this paper is the system noise in (\ref{eq:cnoisy}). SNR for distances between plants ranging from 0 to 2.5 \si{\meter} can observed in Fig. \ref{fig:distancesnr}. 

It can be observed that up to 0.45 \si{\meter}, the SNR value exceeds 40 dB, indicating excellent quality similar to that found in high-fidelity audio systems, digital signals, and high-speed communication systems. As the distance increases to between 0.45 and 1.16 \si{\meter}, the SNR value falls within the range of 30 to 40 dB, which is still suitable for most communications, such as cell phones, TV broadcasts, and clear voice communications. Moving further, within the range of 1.16 to 1.31 \si{\meter}, the SNR value decreases to between 20 and 30 dB. This level is typical of lower-speed communications or slightly degraded systems and is generally considered acceptable quality. When the distance extends from 1.31 to 1.39 \si{\meter}, the SNR value drops to between 10 and 20 dB, which is common in some emergency communications or challenging environments but is regarded as poor quality. Beyond 1.39 \si{\meter}, the SNR value falls below 10 dB, rendering the signal nearly indistinguishable from noise.

Another important metric to investigate in stress communication is the time delay of BVOCs reaching the receiver plant. In this analysis, the time it takes for BVOCs to reach the receiver plant is examined for distances between plants ranging from 0 to 2 meters. When calculating time delays, three approaches are utilized to capture the effect of the relationship between advection and diffusion on time. Time delays are formulated as 
\begin{subequations}\label{eq:timedelays}
\begin{align}
    \tau &= \frac{x_r}{u}, \label{eq:16a}\\
    \tau_1 &= \frac{x_r^2}{D}, \label{eq:16b}\\
    \tau_2 &= \frac{x_r}{2\cdot u} + \frac{x_r^2}{2 \cdot D}, \label{eq:16c}
\end{align}
\end{subequations}
where $\tau$  represent the time delay when advection dominates diffusion, $\tau_1$ represent the time delay when diffusion is the dominant force in the system, and $\tau_2$ represent the time delay when the effects of diffusion and advection are close in magnitude. It can be observed in Fig. \ref{fig:distancedelay} that as advection dominates in the system, time delays are shorter for (\ref{eq:16a}). When time delays are calculated using (\ref{eq:16c}), they are smaller than the values given by (\ref{eq:16b}). Therefore, it can be said that time delays are mostly affected by the intensity of advection compared to diffusion. Further analysis of time delays for different wind speeds will be investigated in the next part.

Lastly, the effect of the amount of released BVOCs from the transmitter plant on communication distance is calculated. Initially, the amount of the message signal, $ (M = 1.1 \cdot 10^{-9}$ \si{\kilo\gram}) was obtained from \cite{copolovici2011volatile}, where the monoterpene emission of \textit{Alnus glutinosa} was recorded under herbivory stress. Other mass values analyzed in this study are multiples of this base value and are used to assess the effect of different masses. As shown in Fig. \ref{fig:distancemass}, keeping other parameters constant, the maximum distance that BVOCs can reach does not change. However, since the message is encoded based on the concentration of BVOCs, it can be seen that the amount of BVOCs is higher for greater distances. Therefore, as expected, increasing the mass of released BVOCs extends the communication distance. Additionally, it can be argued that the quality of communication can improve over greater distances as increasing the mass of released BVOCs enhances the signal strength, resulting in higher SNR values.

\subsubsection{Wind Speed Analysis}
In this analysis, the effect of wind speed on the concentration of BVOCs at the leaves of the receiver plant, $C_{LN}$, is investigated.

Wind speeds are chosen according to the Beaufort Scale, and the main objective of this analysis is to capture a broad range of wind speeds. While $u = 1$ \si{\meter\per\second} represents calm conditions, $u = 100$ \si{\meter\per\second} indicates stormy weather. Hence, while assessing the effects of other parameters in different analyses, wind speed is chosen as $u = 25$ \si{\meter\per\second}, which represents a moderate breeze that may be more applicable to natural conditions.

\begin{figure*}[t]
    \centering
    \begin{subfigure}[b]{0.275\linewidth}
        \centering
        \includegraphics[width=\linewidth]{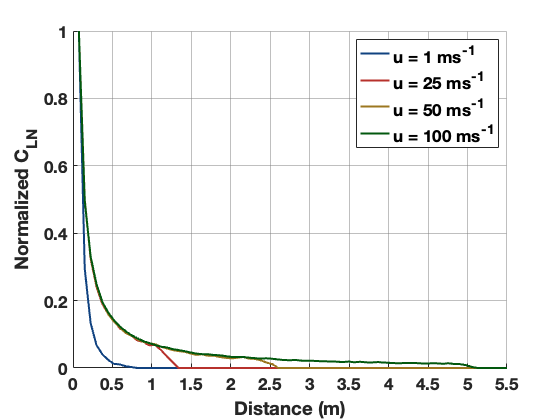}
        \caption{Wind Speed Analysis}
        \label{fig:speed}
    \end{subfigure}
    \hfill
    \begin{subfigure}[b]{0.275\linewidth}
        \centering
        \includegraphics[width=\linewidth]{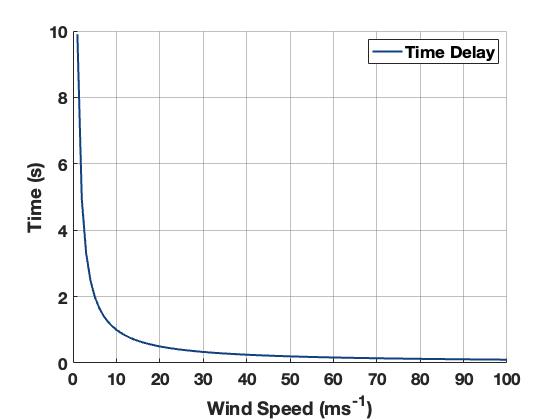}
        \caption{Wind Speed - Delay Analysis}
        \label{fig:speeddelay}
    \end{subfigure}
    \hfill
    \begin{subfigure}[b]{0.275\linewidth}
        \centering
        \includegraphics[width=\linewidth]{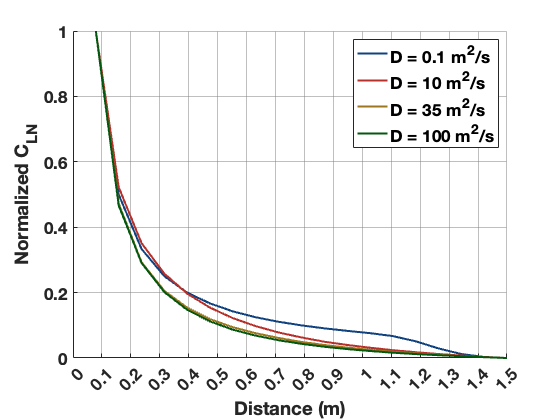}
        \caption{Eddy Diffusivity Analysis  Analysis}
        \label{fig:diffusivity}
    \end{subfigure}

    \vskip\baselineskip

    \begin{subfigure}[b]{0.275\linewidth}
        \centering
        \includegraphics[width=\linewidth]{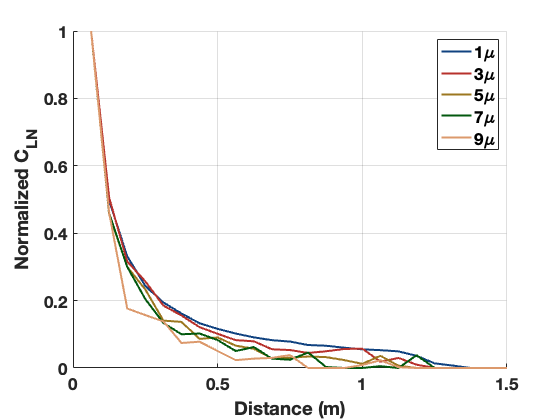}
        \caption{System Noise Intensity Analysis}
        \label{fig:degradation}
    \end{subfigure}
    \hfill
    \begin{subfigure}[b]{0.275\linewidth}
        \centering
        \includegraphics[width=\linewidth]{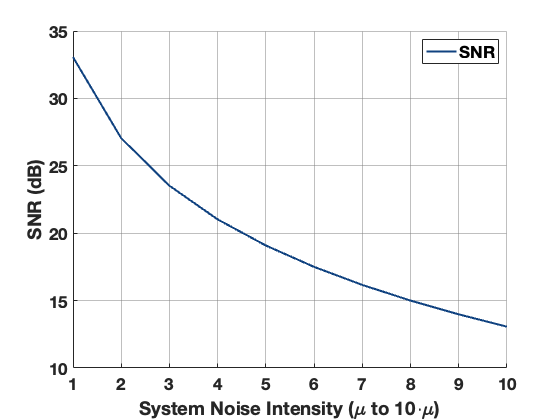}

        \caption{\smaller{System Noise Intensity - SNR Analysis}}
        \label{fig:degradationsnr}
    \end{subfigure}
    \hfill
    \begin{subfigure}[b]{0.275\linewidth}
        \centering
        \includegraphics[width=\linewidth]{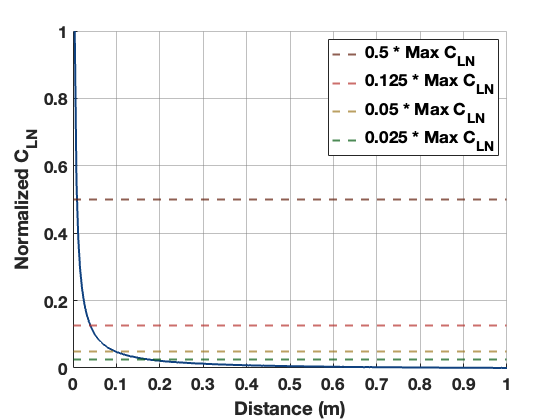}
        \caption{Receiver Threshold Analysis}
        \label{fig:threshold}
    \end{subfigure}
    
    \caption{Wind Speed, Eddy Diffusivity, System Noise Intensity, and Receiver Threshold Analyses}
\end{figure*}

The effects of different wind speeds on $C_{LN}$ can be observed in Fig. \ref{fig:speed}. It can be seen from the figure that, for $u = 100$ \si{\meter\per\second}, BVOCs can reach up to 5 \si{\meter}. On the other hand, for $u = 1$ \si{\meter\per\second}, the maximum extent that BVOCs can reach is around 0.6 \si{\meter}. As expected, an increase in wind speed extends the distance that BVOCs can reach. As wind speeds increase and advection dominates over diffusion, dispersion of BVOCs in directions other than along the wind becomes weaker compared to advection. Hence, as wind speeds increase, the concentration of BVOCs at further distances becomes greater.

Wind speed is an important parameter affecting the time delay in the channel. Therefore, similar to the delay analysis in Section \ref{sec:distanceanalysis}, the delay in the channel should be investigated by comparing different wind speeds. As the effects of wind speed is investigated in this analysis, (\ref{eq:16a}) is employed to calculate time delay. For a receiver plant located at $x_r = 10$ \si{\meter}, the time delay for wind speeds $u = 0 - 100$ \si{\meter\per\second} can be found in Fig. \ref{fig:speeddelay}.

\subsubsection{Eddy Diffusivity Analysis}
In this analysis, the effect of the magnitude of the diffusion coefficient ($D$) on the concentration of BVOCs is investigated. The eddy diffusivity, when constant for other analyses, is chosen as $D  = 0.1 $ \si{\meter\squared\per\second}. This value is obtained from \cite{kulkarni2010atmospheric}, which investigates experimental values of ground-level eddy diffusion coefficients. Choosing different eddy diffusivity coefficients, a wide range of values is selected to observe the effects of diffusion to the greatest extent.

From Fig. \ref{fig:diffusivity}, it can be seen that as the value of $D$ increases, the concentration of BVOCs decreases. This can be explained by the fact that dispersion of BVOCs in directions other than along the wind becomes stronger. However, the maximum distance BVOCs can reach does not differ significantly between different coefficients. Hence, it can be argued that the extent of the distance is predominantly influenced by the wind rather than by diffusion. This also aligns with one of the initial assumptions made in Section \ref{sec:openair}, where turbulence diffusion in the $x$ direction is disregarded.

\subsubsection{System Noise Intensity Analysis}
In this analysis, the effect of the intensity of system noise on the concentration of BVOCs is investigated. Estimating the mean of the system noise is not straightforward. To obtain noise that is well scaled with the value of the signal, the mean of the system noise is chosen as 
\begin{equation}
\begin{aligned} 
\label{eq:degredation}
\mu = \frac{-P_L \cdot A_L}{10 \cdot K_{AW} \cdot M_L} &\frac{M}{8\pi k u} \left( e^{\frac{-(z_r-h)^2-y_r^2}{4k}} + e^{\frac{-(z_r+h)^2-y_r^2}{4k}} \right) \\&  \cdot \left(\operatorname{erf}\left(\frac{u{\tau}_{r} - x_a}{2 \sqrt{k}}\right) + \operatorname{erf}\left(\frac{x_a}{2 \sqrt{k}}\right)\right),
\end{aligned}
\end{equation}
where $x_a$ is calculated for each analysis as half of the maximum distance the BVOCs can reach. Additionally, the standard deviation ($\sigma$) is chosen as $\sigma = \mu / 3$, as 99.7\% of the distribution lies within this region under a Gaussian distribution.

In Fig. \ref{fig:degradation}, it can be seen that as the intensity of system noise increases, the range that BVOCs can reach shortens. Therefore, along the wind, noise in the system can also be defined as another important parameter that affects the maximum distance BVOCs can reach. Additionally, as system noise intensity increases, the fluctuations in concentration become more significant.

Similar to the SNR analysis in Section \ref{sec:distanceanalysis}, another SNR analysis can be conducted to examine the effects of system noise intensity. In this analysis, given a fixed distance of $x_r = 0.75$\si{\meter}, the SNR is calculated by varying the system noise intensity from $\mu$ to $10\cdot\mu$. The effect of system noise intensity on SNR is shown in Fig. \ref{fig:degradationsnr}. It can be seen that, given these noise intensities, the SNR value drops from high quality, which is suitable for most communications, such as cell phones, TV broadcasts, and clear voice communications, to poor quality, which is common in some emergency communications or challenging environments.

\subsubsection{Receiver Threshold Analysis}
\label{sec:thanalysis}
In this analysis, the effect of the threshold value on successful demodulation distance is investigated. As seen in Fig. \ref{fig:threshold}, for a threshold value of $0.5\cdot$ Max($C_{LN}$), the successful demodulation distance is around 0.01 \si{\meter}. As the threshold value decreases, e.g., to $0.025\cdot$ Max($C_{LN}$), this distance can extend up to 0.2 \si{\meter}. However, it should be noted that if the threshold value becomes too small, receiver plants might misclassify constitutive emissions as a message signal, triggering a defense response. This would not be ideal, as entering defense mode alters the behavior of the receiver plant by affecting nutrient uptake and photosynthesis rate \cite{Midzi2022}. On the other hand, if the threshold value is too high, the receiver plant may not capture the message signals sent by stressed transmitter plants, leaving them vulnerable to incoming damage.

\section{Security and Channel Access in Stress Communication}

Up to this point, stress communication between plants has been investigated in terms of the concentration of a single type of BVOC as the message signal, where information is encoded in the concentration. However, as stated in \cite{Midzi2022}, plants also use another modulation technique to prevent competitor plants from preparing for upcoming threats \cite{ueda2012,karban2002,karban2004,yoneya2014,ninkovic2020}. In this modulation technique, plants use a blend of BVOCs as the message signal, where information is encoded in the ratios of individual BVOCs in the blend \cite{ueda2012}. This modulation technique is called Ratio Shift Keying (RSK), which has already been investigated in the context of MC \cite{Kilinc2013,10193791}. In this section, the advantages of this modulation technique will be explained, specifically in terms of enhancing security and enabling multiple users to share communication channels in plant stress signaling.

\subsubsection{Encryption in Stress Communication}
In communication theory, encryption refers to the process of converting information or data into a secure format, so that only authorized parties can access and understand the original message.  Encryption ensures that the content of a message is kept private and cannot be read by unauthorized users during transmission. In nature, by utilizing RSK modulation for message signals, transmitter plants allow only receiver plants of the same species to understand the messages they send \cite{ueda2012}. Therefore, it can be said that plants can achieve encryption in their stress communication using RSK. In this part, the effects of employing RSK on stress communication will be investigated.

Of the five analyses conducted in Section \ref{sec:numanalyses}, only the system noise analysis would lead to changes for RSK. For the other analyses, since the diffusion coefficient is assumed to be the same for both BVOCs, and independent parameters like distance and wind speed affect all BVOC types similarly, the ratios of BVOCs would remain constant across different conditions. However, system noise can impact the BVOCs in the blend differently, leading to a disruption in their ratios and, consequently, a loss of information \cite{ninkovic2020}. Using MATLAB, the effect of system noise on RSK in stress communication is analyzed. The system parameters for this analysis can be found in Table \ref{tab:RSKanalysis}.

\begin{figure*}[t!]
    \centering
    \begin{subfigure}{0.245\linewidth}
        \includegraphics[width=\linewidth]{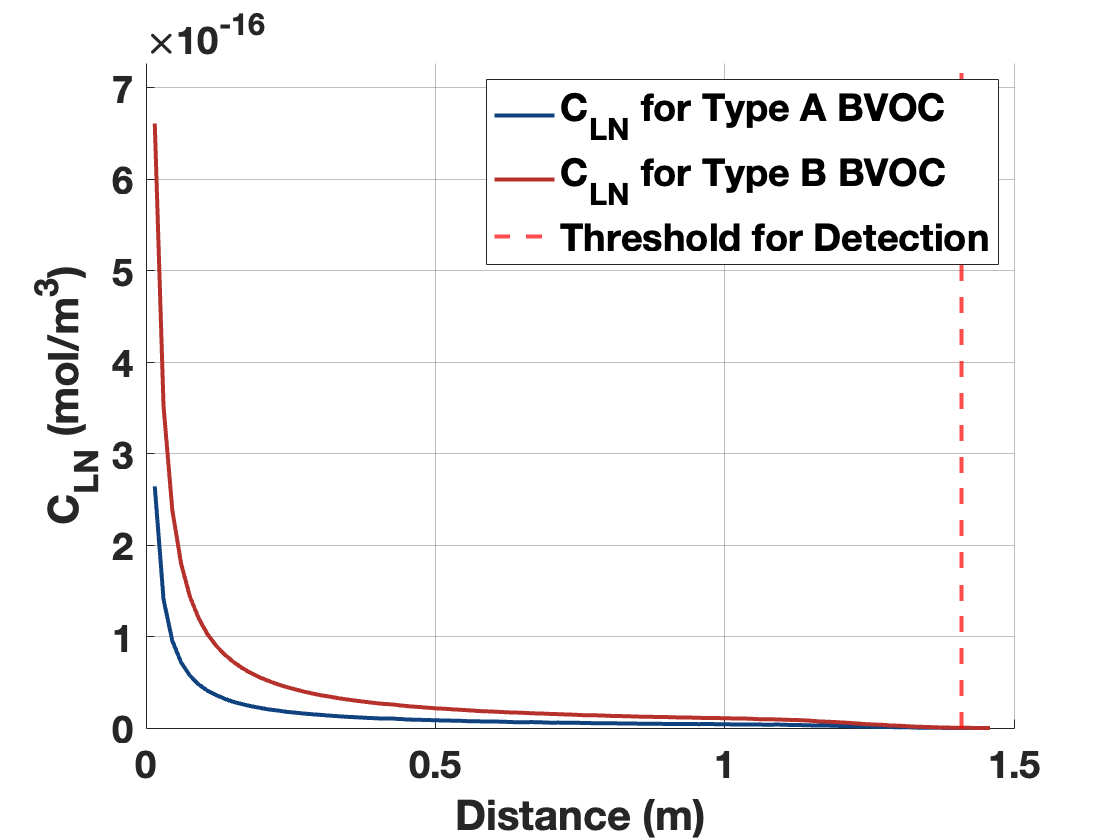}
        \caption{\footnotesize{System Noise Ratio (A/B) : 1/1}}
    \end{subfigure}
    \begin{subfigure}{0.245\linewidth}
        \includegraphics[width=\linewidth]{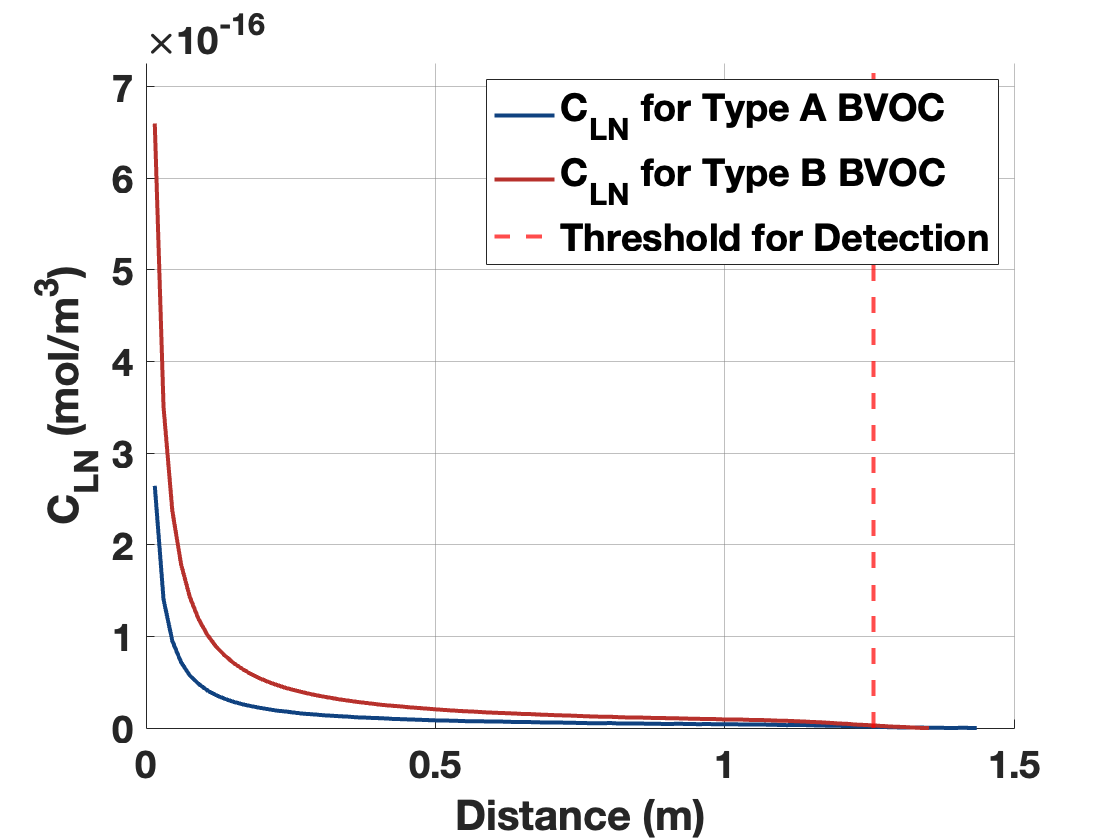}
        \caption{\footnotesize{System Noise Ratio (A/B) : 1/5}}
    \end{subfigure}  
    \begin{subfigure}{0.245\linewidth}
        \includegraphics[width=\linewidth]{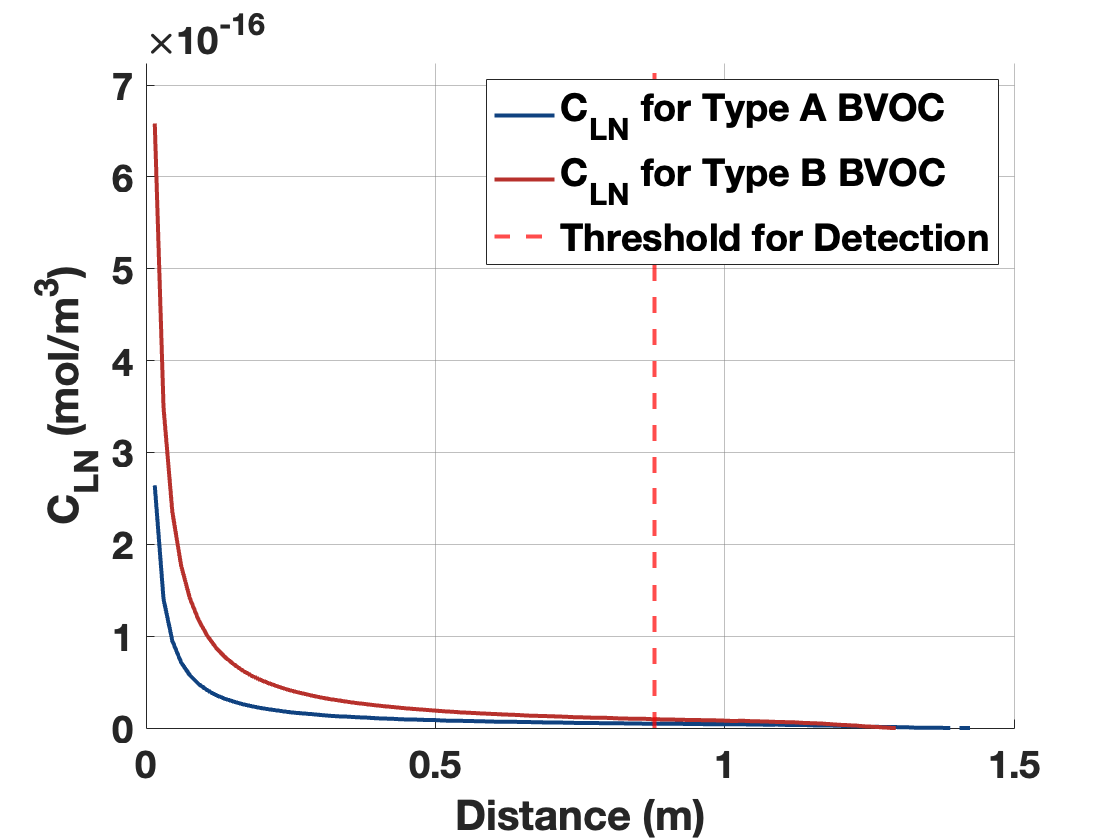}
        \caption{\footnotesize{System Noise Ratio (A/B) : 1/10}}
    \end{subfigure}
    \begin{subfigure}{0.245\linewidth}
        \includegraphics[width=\linewidth]{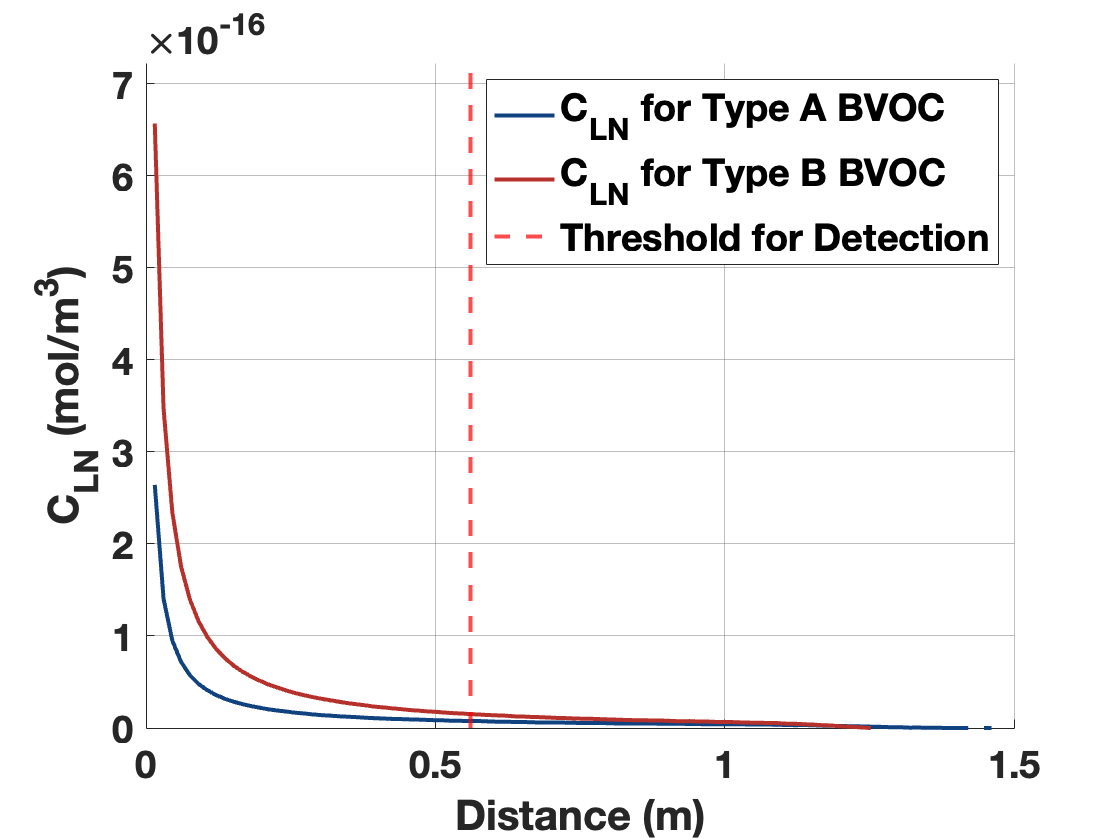}
        \caption{\footnotesize{System Noise Ratio (A/B) : 1/15}}
    \end{subfigure}
    \caption{Ratio Shift Keying Analysis}
    \label{fig:ratioshiftkeyingAnalysis}
\end{figure*}

\begin{table}[h]
  \centering
  \caption{System Parameters for RSK} 
  \label{tab:RSKanalysis}
  \begin{tabular}{|l|l|} 
    \hline
    \textbf{Parameter} & \textbf{Value} \\
    \hline
    $M_A$ & $1.1\cdot10^{-9}$ \si{\kilo\gram} \\ 
    $M_B$ & $2.75\cdot10^{-9}$ \si{\kilo\gram}\\ 
    $P_L$ &  $10^{-8}$ \si{\meter\per\second}\\
    $A_L$ & $25\cdot10^{-4}$ \si{\meter\squared} \\
    $M_L$ & $0.5\cdot10^{-3}$ \si{\kilo\gram} \\
    $K_{AW}$ & 10\\
    $(y_r,z_r)$ & $(0,1) $\\
    $h$ & 1 \si{\meter} \\
    Diffusion Coefficient ($D$) & 0.1 \si{\meter\squared\per\second} \\
    Wind Speed ($u$) & 25 \si{\meter\per\second} \\
    Distance ($x_r$) & 0 - 1.5 \si{\meter}\\
    \hline
  \end{tabular}
\end{table}

In \cite{karlsson2020variability}, it has been shown that different environmental conditions or genotypes can lead to variations in BVOC emissions, including shifts in the blend ratios. Despite these variations, plants can still interpret the altered ratios and respond appropriately to their environment. Given this, the condition for decoding the information for the receiver plant is not set as a strict ratio but rather includes some margin in this analysis. This is a straightforward assumption, as finding BVOCs in nature with an exact ratio is not reasonable. Therefore, in this analysis, the BVOC blend is assumed to consist of two types of BVOCs, and the condition for decoding the information for the receiver plant is set to obtaining the concentration of BVOCs within the ratios of 1:2 and 1:2.5. The encoded information  is accepted as corrupted when the ratio falls below 1:2. The amount of message signal for each BVOC in the blend is denoted as $M_A$ and $M_B$. Parameters $P_L, A_L, M_L, K_{AW}, D, x_r, u$ assumed to be same for both BVOCs.

The system noise analysis for RSK is shown in Fig. \ref{fig:ratioshiftkeyingAnalysis}. In this analysis, the system noise for both type A and type B BVOCs was initially set according to (\ref{eq:degredation}). Subsequently, while keeping the intensity of the system noise for type A BVOC constant, the intensity of the system noise for type B BVOC was increased. By identifying the closest distance at which the ratio of type B to type A BVOC falls below 2, dotted lines were drawn to observe the effective communication distance for plants using RSK methods under different scenarios. As expected, as the intensity of the system noise for type B BVOC increased, the communication distance decreased. Interestingly, the maximum communication distance achieved using RSK was significantly greater than what could be achieved with the lowest threshold in Fig. \ref{fig:threshold}. Therefore, although it might require plants to produce and emit more BVOCs for stress communication, this method could benefit them by both preventing competitor species from eavesdropping and increasing the communication distance over which they can successfully send their messages. Finally, using more than two types of BVOCs in the blend could further extend this communication distance, as the ratio order can be optimized. 

\subsubsection{Multiple Access in Stress Communication}
\begin{figure}[h!]
    \includegraphics[width = \linewidth, height = 4cm]{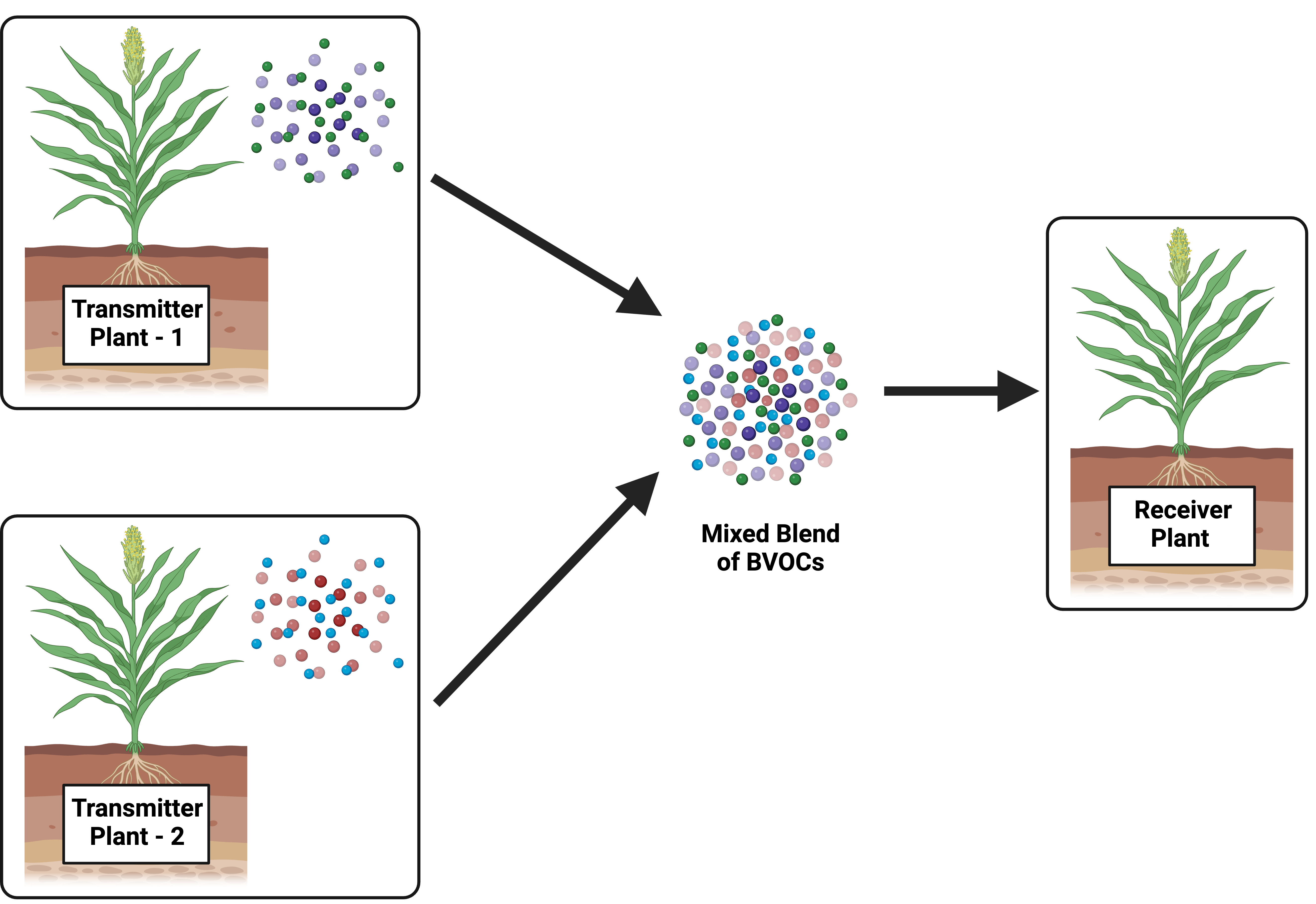 }
    \caption{Scheme of Multiple Access in Stress Communication \cite{fig_3}.}
    \label{fig:MAC}
\end{figure}

In communication theory, multiple access refers to techniques that enable multiple users or devices to share the same communication channel simultaneously. In MC, employing pheromone diversity to establish non-interfering channels can be considered a form of molecular division multiple access (MDMA) \cite{longrangeMC}. This phenomenon can also be observed in stress communication among plants, allowing multiple plants to use the same medium for communication. For the same type of plant species, the composition of volatile plant profiles is altered differently by various stressors \cite{ninkovic2020}. Additionally, the composition of volatile blends depends on plant genotype and differs between various plant species \cite{ninkovic2020}. The overall scheme of multiple access in stress communication can be seen in Fig. \ref{fig:MAC}.

Although RSK enables multiple access in stress communication, the bottlenecks of this phenomenon should also be discussed. First, as mentioned in the previous section, the degradation of BVOCs in volatile blends can cause incorrect demodulation of messages in the system. This can lead to further problems in cases where BVOCs from different blends degrade or chemically interact with each other during transport through the air \cite{ninkovic2020}. Second, if different blends from different plants contain one or more of the same BVOCs, the ratio of the blend arriving at the receiver plant may not be interpretable by the receiver plant.

Analyzing these effects to understand multiple access in stress communication requires further study. However, this analysis can be pursued in future work, as the scope of this paper is to understand the fundamentals of stress communication. We believe that understanding and analyzing stress communication in greater detail is an open research area that could pave the way for sustainable agriculture, plant communication in nature, and preventing agricultural losses.

\section{Conclusion}
In this study, stress communication between plants is investigated. An end-to-end mathematical model is constructed to explain the communication system. Additionally, system components for this communication are defined and explained. To the best of our knowledge, this is the first study that mathematically models this communication through an end-to-end approach. The performance of the system is analyzed numerically using MATLAB under varying wind speeds, diffusion coefficients, system noise intensities, and receiver thresholds.

Using experimental data available in the literature, the theory of continuous regulation of specific gene(s) is validated, providing a model for different stressors across various plants to accurately approximate BVOC emissions.

Another modulation technique observed in plant stress communication, namely Ratio Shift Keying, is explored. This modulation technique is examined under different sytem noise intensities. It has been observed that while this technique allows transmitter plants to encrypt their messages, preventing competitor plants from gaining an advantage, it also enables plants to transmit their messages over greater distances. Additionally, it has also been shown that with this new modulation technique, multiple access in stress communication is achieved.

We believe that understanding stress communication can help us develop realizable OMC systems and improve efficiency in agriculture by deciphering the messages of plants.

\bibliographystyle{IEEEtran}
\bibliography{references.bib}

\begin{IEEEbiography}
    [{\includegraphics[width=1in, height=1.25in, clip, keepaspectratio]{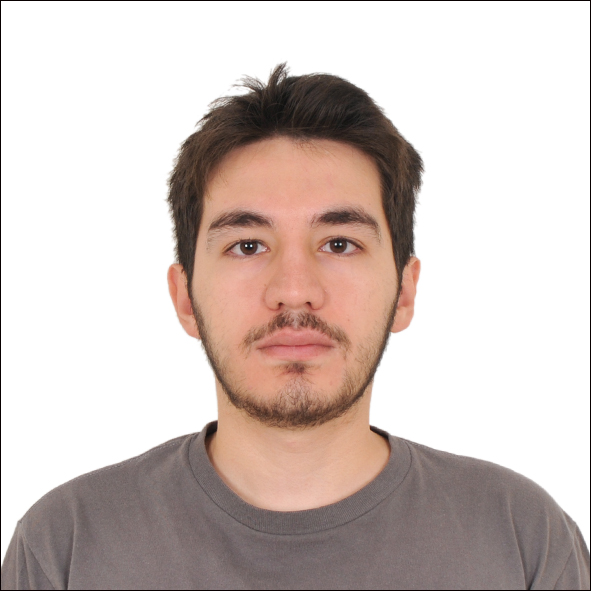}}]{Ahmet Burak Kilic} completed his high school education Bilfen Kayseri High School, Kayseri, Turkey. He is currently a senior student in Electrical and Electronics Engineering with a double major in Business Administration at Koç University, Istanbul, Turkey. He is a research assistant at the Center for neXt-generation Communications (CXC) under the supervision of Prof. Akan.
\end{IEEEbiography}

\begin{IEEEbiography}
[{\includegraphics[width=1in,height=1.25in,clip,keepaspectratio]{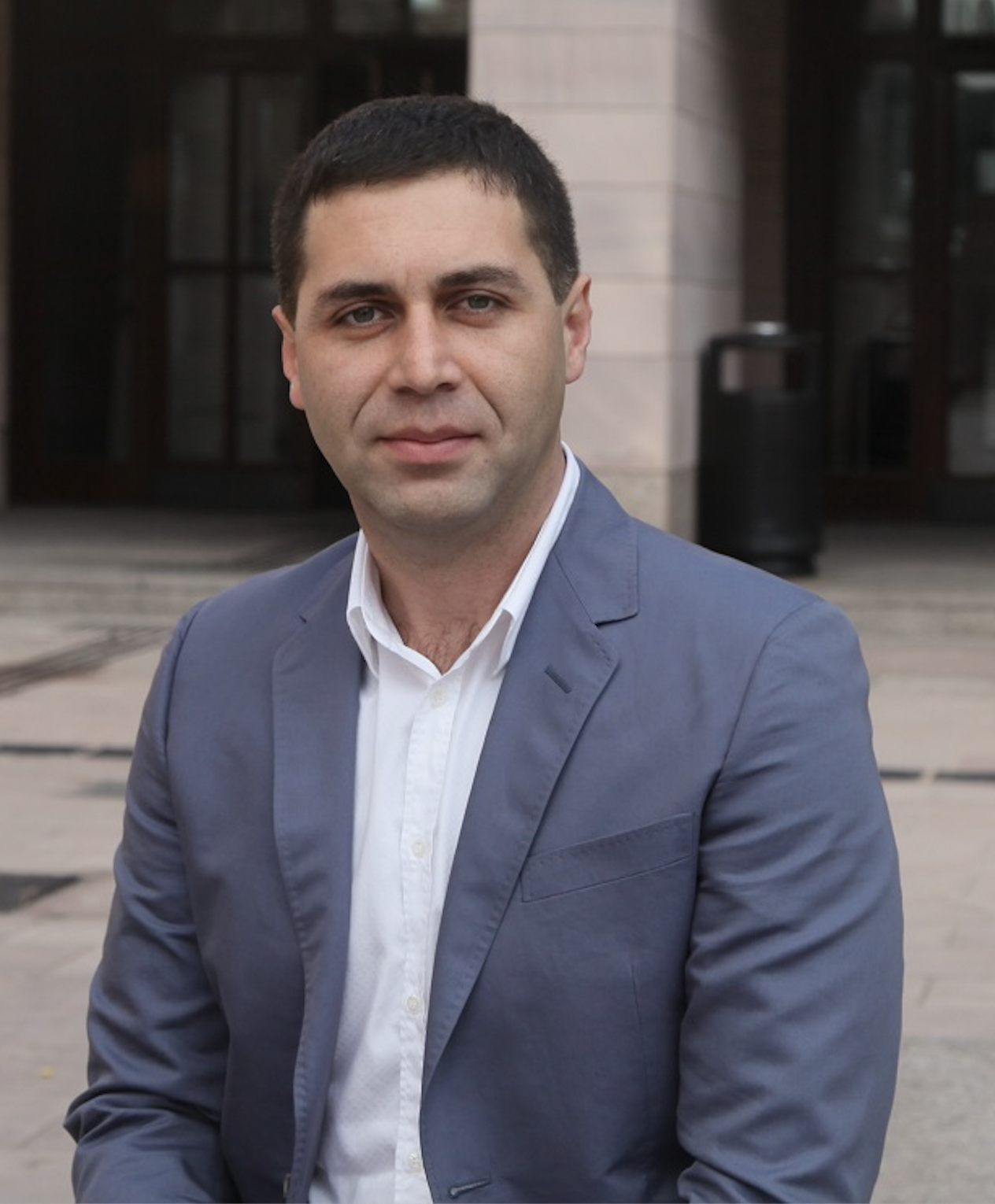}}]{Ozgur B. Akan (Fellow, IEEE)}
received the PhD from the School of Electrical and Computer Engineering Georgia Institute of Technology Atlanta, in 2004. He is currently the Head of Internet of Everything (IoE) Group, with the Department of Engineering, University of Cambridge, UK and the Director of Centre for neXt-generation Communications (CXC), Koç University, Turkey. His research interests include wireless, nano, and molecular communications and Internet of Everything.
\end{IEEEbiography}
\end{document}